\documentclass[onecolumn,11pt]{article}
\usepackage[utf8]{inputenc}

\newcommand{\monthyear}{\ifcase \month \or January\or February\or March\or %
April\or May \or June\or July\or August\or September\or October\or November\or %
December\fi, \number \year}

\newcommand{\dotp}{\ensuremath{\boldsymbol{\cdot}}}

\def\-{\scalebox{0.4}[1.0]{\(-\)}}

\newcommand{\mathcalOM}[1]{\ensuremath{\mathcal{O}(M^{#1})}}

\newcommand{\ord}[2]{\ensuremath{#1^{(#2)}}}

\newcommand{\uuline}[1]{\underline{\underline{#1}}}

\usepackage{verbatim}

\usepackage{setspace}

\usepackage{amsmath}
\usepackage{amssymb}

\usepackage{mathtools}

\usepackage[hidelinks]{hyperref}

\usepackage[doi=false,url=false,isbn=false,giveninits=true]{biblatex}
\AtEveryBibitem{%
    \clearfield{note}%
    \clearlist{language}%
    \clearfield{month}%
    \clearfield{publisher}%
    \clearfield{pages}%
}
\renewbibmacro{in:}{}
\usepackage[a4paper, margin=1.5cm]{geometry}

\usepackage[textsize=tiny]{todonotes}

\usepackage{cancel}

\usepackage{stmaryrd}

\usepackage{graphicx}
\usepackage{graphics}
\usepackage{epstopdf}

\usepackage{subcaption}

\usepackage{svg}

\usepackage{multirow}
\usepackage{tabularx}

\usepackage{appendix}
\usepackage{xcolor}

\usepackage[perpage,symbol*]{footmisc}
\DefineFNsymbols*{lamportnostar}[math]{\dagger\ddagger\S\P\|{\dagger\dagger}{\ddagger\ddagger}}
\setfnsymbol{lamportnostar}

\usepackage{lineno}

\begin{document}

\title{Artificial diffusion for convective and acoustic low Mach number flows II: Application to Liou-Steffen, Zha-Bilgen and Toro-Vasquez convection-pressure flux splittings}
\author{Joshua Hope-Collins\footnote{Oxford Thermofluids Institute, University of Oxford, Osney Mead, Oxford, OX2 0ES, England, United Kingdom} \footnote{Department of Mathematics, Imperial College London, London, SW7 2AZ, England, United Kingdom} \footnote{Corresponding author. \textit{Email address}: joshua.hope-collins13@imperial.ac.uk}\\
Luca di Mare$^{\dagger}$}
\date{\monthyear}

\maketitle
\abstract{

\noindent Liou-Steffen splitting (AUSM) schemes are popular for low Mach number simulations, however, like many numerical schemes for compressible flow they require careful modification to accurately resolve convective features in this regime.
Previous analyses of these schemes usually focus only on a single discrete scheme at the convective limit, only considering flow with acoustic effects empirically, if at all.
In our recent paper \emph{Hope-Collins \& di Mare, 2023} we derived constraints on the artificial diffusion scaling of low Mach number schemes for flows both with and without acoustic effects, and applied this analysis to Roe-type finite-volume schemes.
In this paper we form approximate diffusion matrices for the Liou-Steffen splitting, as well as the closely related Zha-Bilgen and Toro-Vasquez splittings.
We use the constraints found in \emph{Hope-Collins \& di Mare, 2023} to derive and analyse the required scaling of each splitting at low Mach number.
By transforming the diffusion matrices to the entropy variables we can identify erroneous diffusion terms compared to the ideal form used in \emph{Hope-Collins \& di Mare, 2023}.
These terms vanish asymptotically for the Liou-Steffen splitting, but result in spurious entropy generation for the Zha-Bilgen and Toro-Vasquez splittings unless a particular form of the interface pressure is used.
Numerical examples for acoustic and convective flow verify the results of the analysis, and show the importance of considering the resolution of the entropy field when assessing schemes of this type.\\


\noindent Highlights:
\begin{itemize}
    \item Formulation of approximate diffusion matrices for convection-pressure flux splittings
    \item Identification of required asymptotic scaling of diffusion coefficients
    \item Considers both convective and acoustic flow features
    \item Resolution of the entropy field is particularly sensitive to the choice of splitting
    \item Numerical results show excellent agreement with analysis
\end{itemize}
}

\newpage
\tableofcontents
\newpage

\graphicspath{{./figures/}}

\section*{Introduction}\label{sec:intro}

Low Mach number flows occur in many applications of engineering and scientific interest, and the design of numerical schemes that can accurately simulate these flows has been an active research area for several decades \cite{turkel_review_1993,guillard_chapter_2017}.
It is well-known that collocated density-based schemes for compressible flow may fail to accurately resolve convective flow features at low Mach number due to the incorrect scaling of the artificial diffusion terms in the asymptotic limit $M\to0$ \cite{guillard_behaviour_1999,turkel_preconditioning_1999}.
A great many schemes which overcome this failing have been developed, including central-difference schemes \cite{turkel_preconditioned_1987,turkel_review_1993,venkateswaran_artficial_2003} and flux-difference splitting schemes \cite{guillard_behaviour_1999,guillard_behavior_2004,thornber_improved_2008,rieper_low-mach_2011}, as well as convection-pressure flux-vector splitting schemes \cite{liou_numerical_1999,liou_sequel_2006,shima_parameter-free_2011} which are the focus of the current work.

Convection-pressure flux-vector splittings separate the physical flux of the Euler equations into two terms: one describing the transport of some set of convected quantities, and the other containing the contribution of the pressure.
This is in contrast with other flux-vector splitting methods where the flux vector is split into a forward travelling wave and a backward travelling wave \cite{steger_flux_1981,van_leer_flux-vector_1982}.
A numerical flux function can be constructed from a convection-pressure flux-vector splitting by finding suitable discretisations for each term, and fluxes of this type are attractive for a number of reasons.
Separating the flux allows different modelling approaches to be used for each term, which is especially useful at low Mach number where the velocity and pressure play increasingly distinct roles \cite{muller_low_1999}.
Splitting in this manner also gives a balance of accuracy and computational efficiency: intermediate contact waves can be accurately resolved unlike standard flux-vector splitting schemes, but the computational expense scales with the size $n$ of the system being solved instead of $n^2$ for flux-difference splittings, which is especially useful for multi-component flows \cite{liou_new_1993,liou_sequel_1996}.

The most commonly used convection-pressure flux-vector splitting is the Liou-Steffen splitting, named advection upstream splitting method (AUSM) in the original papers from Liou \& Steffen 1993 and Liou 1996 \cite{liou_new_1993,liou_sequel_1996}.
This splitting has been widely applied, including to low Mach number \cite{liou_sequel_2006}, hypersonic \cite{kitamura_reduced_2016}, multicomponent and multiphase \cite{paillere_extension_2003,kitamura_extension_2014} flows.
The earliest low Mach number schemes were inspired by the ideas of low Mach number preconditioning \cite{edwards_low-diffusion_1998,liou_numerical_1999,edwards_towards_2001}, which can be combined with flux-difference splitting schemes to overcome the accuracy problem at low Mach number \cite{turkel_preconditioning_1999,guillard_behaviour_1999}.
Liou 2006 \cite{liou_sequel_2006} integrated these ideas into the AUSM$^+$-up for all speeds scheme, which is still in widespread use today.
Of the more recent Liou-Steffen splitting schemes suitable for low Mach number, SLAU and its descendants are particularly popular \cite{shima_parameter-free_2011,kitamura_towards_2013,kitamura_reduced_2016}.
Two other convection-pressure flux-vector splittings are the Zha-Bilgen \cite{zha_numerical_1993} and Toro-Vasquez \cite{toro_flux_2012} splittings, which differ from the Liou-Steffen splitting only in the treatment of the energy equation.
The Zha-Bilgen splitting was derived by splitting the exact flux Jacobian into convective and acoustic components in \cite{schiff_numerical_1980,steger_flux_1981}.
Zha \& Bilgen 1993 \cite{zha_numerical_1993} developed a numerical scheme for this splitting, and as such it is usually referred to by their names.
The Toro-Vasquez splitting was proposed by Toro \& Vasquez 2012 \cite{toro_flux_2012} in the context of developing a general framework for the discretisation of convection-pressure flux-vector splittings.
Over the last decade, the Zha-Bilgen and Toro-Vasquez splittings have gained some attention for low Mach number flow \cite{qu_new_2018,chen_novel_2018,iampietro_low-diffusion_2020,sun_robust_2017,lin_density_2018,chen_low-diffusion_2021}, and it has been shown that accurate schemes for this regime can be produced from these splittings.
We cover more detail on the literature for low Mach number schemes using convection-pressure flux-vector splittings in section \ref{sec:existing-schemes}.\\

In our previous paper \emph{``Artificial diffusion for convective and acoustic low Mach number flows I: Analysis of the modified equations, and application to Roe-type schemes''} \cite{hope-collins_artificial_2022} we analysed the form and behaviour of artificial diffusion in numerical schemes at low Mach number.
Three diffusion scalings naturally arise - one for purely convective flow, one for purely acoustic flow, and one for flow with either or both convective and acoustic features.
Our approach used the modified equations, which make the artificial diffusion terms of the discrete numerical scheme explicit in the continuous PDE and enable the behaviour of these terms to be studied independently of the specific discretisation they arose from.
The modified equations have previously been used to study low Mach number numerical schemes in \cite{turkel_preconditioning_1999,dellacherie_analysis_2010}.
By applying this analysis to Roe-type schemes, we showed that many existing low Mach number schemes from the literature adhere to one of the three diffusion scalings, and display the expected behaviour.
While the behaviours of these schemes have been shown and studied previously, this approach demonstrated that many properties of low Mach number schemes can be predicted and understood using the continuous equations, and that considering both convective and acoustic flow is necessary to provide a complete picture of this behaviour.

In this paper, we apply this approach to investigate the behaviour of convection-pressure flux-vector splitting schemes at low Mach number, specifically the Liou-Steffen, Zha-Bilgen, and Toro-Vasquez splittings.
By forming explicit diffusion matrices for each splitting, we can clearly see the differences in the forms of the diffusion, and by transforming to the entropy variables we can make direct comparison to the results of \cite{hope-collins_artificial_2022}.
Many previous studies of low Mach number numerical schemes have focused on Roe-type schemes - as we did in \cite{hope-collins_artificial_2022} - and the findings are extended to convection-pressure flux-vector splitting schemes by identifying the diffusion terms that these schemes have in common with the Roe-type schemes \cite{dellacherie_analysis_2010,sachdev_improved_2012,li_mechanism_2013}.
While we also do this here, working in the entropy variables means that we can additionally show the effect of the diffusion terms which are different between the convection-pressure flux-vector splitting schemes and the Roe-type schemes.
The particular form of these terms turns out to be crucial to the success of these schemes at low Mach number.

The structure of the remainder of the paper is as follows.
In section \ref{sec:lowmach-euler} we very briefly cover the necessary background for the low Mach number limits of the Euler equations, and in section \ref{sec:artificial-diffusion} we recap some of the key findings of \cite{hope-collins_artificial_2022} which will be used later.
Section \ref{sec:cp-splittings} contains the main content of this study.
The approximate diffusion matrix for each splitting is shown in the conserved variables, then transformed to the entropy variables.
The form in the entropy variables is discussed, and the limiting form of the modified equations for each splitting at both the convective and acoustic limits are found.
In section \ref{sec:existing-schemes} we look at a number of existing schemes from the literature, compare them to the generic forms used in section \ref{sec:cp-splittings}, and identify which of the three diffusion scalings they use.
The results of several numerical examples are shown in section \ref{sec:numerical-examples} and display excellent agreement with the analysis of section \ref{sec:cp-splittings} for all three splittings for both convective and acoustic flow.

\section{Low Mach number limits of the Euler equations}\label{sec:lowmach-euler}

Here we give a brief recap of the necessary background for the low Mach number asymptotic limits of the Euler equations.
More details are given in \cite{muller_low_1999,dellacherie_analysis_2010,hope-collins_artificial_2022}.
We use the non-dimensionalisation:
\begin{equation}\label{eq:non-dimensionalisation}
\begin{gathered}
    \rho = \frac{\tilde{\rho}}{\rho_{\infty}},
    \quad
    \underline{u} = \frac{\underline{\tilde{u}}}{u_{\infty}},
    \quad
    \underline{x} = \frac{\underline{\tilde{x}}}{L_{\infty}},
    \quad
    E = \frac{\tilde{E}}{p_{\infty}/\rho_{\infty}},
    \quad
    H = \frac{\tilde{H}}{p_{\infty}/\rho_{\infty}},
    \quad
    p = \frac{\tilde{p}}{p_{\infty}},
    \quad
    t = \frac{\tilde{t}}{L_{\infty}/u_{\infty}}
\end{gathered}
\end{equation}
where tildes indicate local dimensional quantities, $\infty$ indicates reference dimensional quantities, and underlining indicates vector-valued quantities.
$E$ and $H$ are the specific total energy and enthalpy respectively.
Using this non-dimensionalisation, the Euler equations in the entropy variables $d\underline{w}=(dp,du,dv,ds)^T$ where $ds=dp-a^2d\rho$ are:
\begin{equation} \label{eq:euler}
\begin{aligned}
    \partial_{t} p + \underline{u}\dotp\nabla p + \gamma p\nabla\dotp\underline{u} = 0 \\
    \rho\partial_{t} \underline{u} + M^{\-2}\nabla p + \rho\underline{u}\dotp\nabla\underline{u} = 0 \\
    \partial_{t} s + \underline{u}\dotp\nabla s = 0
\end{aligned}
\end{equation}

where $M = \frac{\sqrt{\gamma}u_{\infty}}{a_{\infty}}$ is the reference Mach number.
Two single scale limits will be shown here for purely convective and purely acoustic flow.
A multiple time (space) scale, single space (time) scale limit which contains both convective and acoustic flow features also exists, see \cite{hope-collins_artificial_2022} and \cite{muller_low_1999,bruel_low_2019} for more detail.
To find the convective limit, all variables in (\ref{eq:euler}) are expanded as power series of the Mach number:
\begin{equation} \label{eq:power_expansion}
    \psi(\underline{x},t,M) =     \ord{\psi}{0} (\underline{x},t)
                             + M  \ord{\psi}{1} (\underline{x},t)
                             + M^2\ord{\psi}{2} (\underline{x},t)
                             + \mathcalOM{3}
\end{equation}
Grouping terms of equal order, we find the following relations for the pressure and velocity:
\begin{equation} \label{eq:convective_timescale}
\begin{aligned}
    \nabla\,\ord{p}{0,1} = 0 \\
    \ord{\rho}{0}\partial_{t} \ord{\underline{u}}{0} + \nabla \ord{p}{2} + \ord{\underline{\rho u}}{0}\dotp\nabla\ord{\underline{u}}{0} = 0 \\
    d_{t} \ord{p}{0,1} + \gamma\ord{(p\nabla\dotp\underline{u})}{0,1} = 0
\end{aligned}
\end{equation}
The pressure has only $\mathcalOM{2}$ spatial variations, and at steady-state the divergence of the zeroth and first order velocities vanish, with the system (\ref{eq:convective_timescale}) approaching the incompressible regime as $M\to0$.
The same relation is found for every order of the entropy equation:
\begin{equation}
    \partial_{t}\ord{s}{n} + \ord{(\underline{u}\dotp\nabla s)}{n} = 0
\end{equation}

The acoustic limit can be found by replacing the convective reference timescale $L_{\infty}/u_{\infty}$ with the acoustic reference timescale $L_{\infty}/a_{\infty}$ in the non-dimensionalisation (\ref{eq:non-dimensionalisation}).
The acoustic non-dimensional time is then $\tau=a_{\infty}\tilde{t}/L_{\infty}=t/M$.
Repeating the expansion with this timescale we find:
\begin{equation} \label{eq:acoustic_timescale}
\begin{aligned}
    \nabla\,\ord{p}{0} = \partial_{\tau}\ord{p}{0} = 0 \\
    \ord{\rho}{0}\partial_{\tau}\ord{\underline{u}}{0} + \nabla \ord{p}{1} = 0 \\
    d_{\tau} \ord{p}{1} + \gamma\ord{p}{0}\nabla\dotp\ord{\underline{u}}{0} = 0
\end{aligned}
\end{equation}
The pressure now has $\mathcalOM{1}$ spatial variations, the zeroth order velocity divergence is non-zero, and the system (\ref{eq:acoustic_timescale}) approaches the equations for low Mach number acoustics.
The leading order relation from the entropy equation is
\begin{equation}
    \partial_{\tau}\ord{s}{0} = 0
\end{equation}
which shows that the (leading order) entropy does not vary on the acoustic timescale.
The asymptotic scaling of quantities at each of the two regimes are shown in table \ref{tab:lowmach_scaling}, and will be used later to find the limit equations for different numerical schemes in each regime.

\begin{table}
\def\-{\scalebox{0.4}[1.0]{\(-\)}}
\begin{center}
\begin{tabular}{|c|c|c|} \hline
                    & Convective      & Acoustic         \\ \hline
     $p$            & \multicolumn{2}{c|}{$\ord{p}{0}   \sim\mathcalOM{0}$} \\ \hline
     $u$            & \multicolumn{2}{c|}{$\ord{u}{0}   \sim\mathcalOM{0}$} \\ \hline
     $\rho$         & \multicolumn{2}{c|}{$\ord{\rho}{0}\sim\mathcalOM{0}$} \\ \hline
     $\partial_x p$ & $\partial_x\ord{p}{2}\sim\mathcalOM{2}$ & $\partial_x\ord{p}{1}\sim\mathcalOM{}$  \\ \hline
     $\partial_x u$ & \multicolumn{2}{c|}{$\partial_x\ord{u}{0}\sim\mathcalOM{0}$} \\ \hline
     $\partial_x v$ & \multicolumn{2}{c|}{$\partial_x\ord{v}{0}\sim\mathcalOM{0}$} \\ \hline
     $\partial_x s$ & \multicolumn{2}{c|}{$\partial_x\ord{s}{0}\sim\mathcalOM{0}$} \\ \hline
     $\partial_t p$ & $\partial_t\ord{p}{0}\sim\mathcalOM{0}$ & $\partial_{\tau}\ord{p}{1}\sim\mathcalOM{0}$  \\ \hline
     $\partial_t u$ & $\partial_t\ord{u}{0}\sim\mathcalOM{0}$ & $\partial_{\tau}\ord{u}{0}\sim\mathcalOM{\-1}$ \\ \hline
     $\partial_t v$ & \multicolumn{2}{c|}{$\partial_t\ord{v}{0}\sim\mathcalOM{0}$} \\ \hline
     $\partial_t s$ & \multicolumn{2}{c|}{$\partial_t\ord{s}{0}\sim\mathcalOM{0}$} \\ \hline
\end{tabular}
\caption{Low Mach number scaling of the various terms in equations (\ref{eq:convective_timescale}), (\ref{eq:acoustic_timescale}) for convective or acoustic variations}
\label{tab:lowmach_scaling}
\end{center}
\end{table}

\section{Artificial diffusion for low Mach number schemes}\label{sec:artificial-diffusion}

In \cite{hope-collins_artificial_2022}, the $x$-split form of the modified equations in the entropy variables was used to determine the appropriate scaling of the artificial diffusion at low Mach number.
Although convective and acoustic flow features can only properly be distinguished with a truly multi-dimensional analysis \cite{dellacherie_analysis_2010,barsukow_truly_2021}, many finite-volume/difference schemes use a dimension splitting procedure, and many low Mach number behaviours of these schemes can be demonstrated with the 1D-split equations - with the understanding that the final scheme will be the sum over all dimensions.
The modified equations make the artificial diffusion terms which are inherent in many numerical schemes explicit in the continuous equations \cite{warming_modified_1974}, which allows analysis of the behaviour of these terms without restriction to a specific discrete form.
The modified equations in the symmetry variables were used in \cite{turkel_preconditioning_1999} to demonstrate the behaviour of the artificial diffusion of the standard and preconditioned Roe scheme, and the modified equations of the barotropic Euler equations were used in \cite{dellacherie_analysis_2010} to investigate the requirements for accurate schemes at the convective limit.
The entropy variables are useful at low Mach number because in the 1D-split form the vorticity (transverse-velocity) and entropy equations are equivalent to scalar advection equations, which require only the usual upwind diffusion using the convective velocity scale.
This reduces the degrees of freedom we need to consider to only those in the pressure and (normal-)velocity equations, which are the distinguishing feature between the different low Mach number limits and schemes.

In the first part of this section, the form of the artificial diffusion in the entropy variables used in \cite{hope-collins_artificial_2022} will be introduced, and the required scaling at each of the three low Mach number limits will be recalled and briefly discussed.
In the second part of the section, the form of the artificial diffusion transformed to the conserved variables will be presented, which will be useful later for comparison with the diffusion of the convection-pressure flux-vector splitting schemes.

\subsection{Entropy variables form}
The form of the modified equations in the non-dimensional entropy variables used in \cite{hope-collins_artificial_2022} to investigate the behaviour of the artificial diffusion is:\footnote{The form written here is equivalent to equation (14) with the diffusion coefficients from equation (41) in \cite{hope-collins_artificial_2022}.}
\begin{equation} \label{eq:euler_modified}
    \begin{aligned}
        \partial_t p +           u\partial_x p +\gamma p\partial_x u & = \phantom{\rho}\mu_u|u|\partial_{xx}p + M^{\-2}\frac{\gamma p}{\rho|v|}\mu_{11}\partial_{xx}p \pm \gamma p\mu_{12}\partial_{xx}u \\
        \rho\partial_t u + M^{\-2}\partial_x p +  \rho u\partial_x u & = \rho\mu_u|u|\partial_{xx}u \phantom{\frac{\gamma p}{\rho|v|}}\pm M^{\-2}\mu_{21}\partial_{xx}p + \rho|v|\mu_{22}\partial_{xx}u \\
        \partial_t v                           +       u\partial_x v & = \phantom{\rho}\mu_u|u|\partial_{xx}v \\
        \partial_t s                           +       u\partial_x s & = \phantom{\rho}\mu_u|u|\partial_{xx}s
    \end{aligned}
\end{equation}
The terms with the $\mu_u$ diffusion coefficient are the convective upwinding, so $\mu_u\sim\mathcalOM{0}$, and $|v|$ is some velocity scale.
The diffusion terms with the $\mu_{ij}$, $i,j=1,2$ coefficients are elements of the artificial diffusion Jacobian $\uuline{A}$.
The form of the Jacobian elements used here was chosen because it simplifies the diffusion coefficients once we transform to the conserved variables below.
The precise form of $\mu_{ij}$, $i,j=1,2$ is particular to each discrete scheme.
The sign of the off-diagonal terms may be negative so long as the diffusion Jacobian remains positive (semi-)definite.
See \cite{hope-collins_artificial_2022} for a discussion of the requirements on the form of the off-diagonal diffusion components.

In \cite{hope-collins_artificial_2022} the method of Turkel 1999 \cite{turkel_preconditioning_1999} was used to find the limiting forms of (\ref{eq:euler_modified}) for each of the convective and acoustic limits by enforcing the variations in table \ref{tab:lowmach_scaling}.
The Mach number scaling of the coefficients $\mu_{ij}$ was then chosen for each of the two limits such that the artificial diffusion terms had the same Mach number scaling as the dominant physical terms on the left hand side.
This ensures that the artificial diffusion terms do not vanish asymptotically - which would result in loss of stability - nor do they dominate asymptotically - which would result in loss of accuracy.
The $M^{\-2}$ coefficients in $A_{11}$ and $A_{21}$ are included so that at the convective limit all $\mu_{ij}$ are independent of the Mach number, as is appropriate for the incompressible regime, while keeping the correct scaling for the full Jacobian elements $A_{ij}$.
The convective and acoustic diffusion scalings were then combined into a mixed diffusion scaling which is suitable for both convective and acoustic flow features.
The scaling of the coefficients for the convective ($\uuline{A}^c$), acoustic ($\uuline{A}^a$) and mixed ($\uuline{A}^m)$ diffusion Jacobians are shown in table \ref{tab:diffusion-scaling}.
Each of these three diffusion scalings has its own strengths and weaknesses.
A brief description is given here, with more detail given in \cite{hope-collins_artificial_2022} and references therein.

\begin{table}
\def\-{\scalebox{0.4}[1.0]{\(-\)}}
    \centering
    \renewcommand{\arraystretch}{1.5}
    \begin{tabular}{|c|c|c|c|} \hline
        & $\uuline{A}^c$ & $\uuline{A}^a$ & $\uuline{A}^m$ \\ \hline
        {\renewcommand{\arraystretch}{1} $
        \begin{matrix} \begin{pmatrix}
        A_{11} & A_{12} \\
        A_{21} & A_{22}
        \end{pmatrix} \end{matrix}$ } &
        {\renewcommand{\arraystretch}{1} $\mathcal{O}
        \begin{matrix} \begin{pmatrix}
        M^{\-2} & M^{0} \\
        M^{\-2} & M^{0}
        \end{pmatrix} \end{matrix}$} &
        {\renewcommand{\arraystretch}{1} $\mathcal{O}
        \begin{matrix} \begin{pmatrix}
        M^{\-1} & M^{0} \\
        M^{\-2} & M^{\-1}
        \end{pmatrix} \end{matrix}$} &
        {\renewcommand{\arraystretch}{1} $\mathcal{O}
        \begin{matrix} \begin{pmatrix}
        M^{\-1} & M^{0} \\
        M^{\-2} & M^{0}
        \end{pmatrix} \end{matrix}$} \\ \hline
        {\renewcommand{\arraystretch}{1} $
        \begin{matrix} \begin{pmatrix}
        \mu_{11} & \mu_{12} \\
        \mu_{21} & \mu_{22}
        \end{pmatrix} \end{matrix}$ } &
        {\renewcommand{\arraystretch}{1} $\mathcal{O}
        \begin{matrix} \begin{pmatrix}
        M^{0} & M^{0} \\
        M^{0} & M^{0}
        \end{pmatrix} \end{matrix}$} &
        {\renewcommand{\arraystretch}{1} $\mathcal{O}
        \begin{matrix} \begin{pmatrix}
        M^{} & M^{0} \\
        M^{0} & M^{\-1}
        \end{pmatrix} \end{matrix}$} &
        {\renewcommand{\arraystretch}{1} $\mathcal{O}
        \begin{matrix} \begin{pmatrix}
        M^{} & M^{0} \\
        M^{0} & M^{0}
        \end{pmatrix} \end{matrix}$} \\ \hline
        Suitable for convective flow & Yes  & No & Yes$^{1}$  \\ \hline
        Suitable for acoustic flow   & No & Yes  & Yes$^{2}$  \\ \hline
        Timestep limit & \mathcalOM{2} & \mathcalOM{1} & \mathcalOM{1} \\ \hline
    \end{tabular}
    \caption{Details of the convective $\uuline{A}^c$, acoustic $\uuline{A}^a$, and mixed $\uuline{A}^m$ diffusion scalings. The mixed scheme is susceptible to $^{1}$pressure chequerboard instabilities at the convective limit and $^{2}$velocity instabilities at the acoustic limit.}
    \label{tab:diffusion-scaling}
\end{table}

The convective diffusion scaling has a large $\mathcalOM{\-2}$ pressure diffusion coefficient $A_{11}$ to compensate for the $\mathcalOM{2}$ scaling of $\partial_{x}p$ at the convective limit.
This term prevents pressure chequer-board instabilities, which result from the same pressure-velocity decoupling that occurs with collocated schemes for incompressible flow.
This increased pressure diffusion also damps out any acoustic features in the flow, making the convective scheme unsuitable for the acoustic or mixed limits.
The pressure diffusion also increases the spectral radius of the diffusion matrix by a factor of $\mathcalOM{\-1}$ compared to the physical flux Jacobian and the other two diffusion matrix scalings.
This means that an implicit time integration strategy is required for time-accurate simulations, as the timestep restriction scale as $\Delta t\sim\mathcalOM{2}$ instead of the usual $\Delta t\sim\mathcalOM{}$ for stability of the explicit scheme.

The acoustic diffusion scaling has a smaller pressure diffusion coefficient, so can resolve acoustic flow features.
However, the velocity diffusion $A_{22}$ is a factor of $\mathcalOM{\-1}$ larger than for the convective scaling, which destroys the accuracy for convective flow features and makes this scaling unsuitable for the convective or mixed limits.
Most schemes for compressible flow designed for higher Mach number regimes approach this scaling at low Mach number, and hence are inaccurate at low Mach number unless at least this term is modified.

The mixed scaling has neither the $\mathcalOM{\-2}$ pressure diffusion of the convective diffusion nor the $\mathcalOM{\-1}$ velocity diffusion of the acoustic scaling.
As such it does not overdamp acoustic nor convective flow features and is accurate for the convective, acoustic, and mixed limits, although this flexibility does not come without cost.
The pressure diffusion vanishes asymptotically at the purely convective limit, meaning that this scheme is susceptible to pressure chequerboard instabilities.
The velocity diffusion vanishes asymptotically at the purely acoustic limit, meaning that there is an undamped grid mode which can be excited by non-linearities, initial or boundary conditions, poor mesh quality, and acoustic discontinuities such as shockwaves.

\subsection{Conserved variables form}

Transforming the modified equations in the non-dimensional entropy variables (\ref{eq:euler_modified}) to the dimensional conserved variables, we can construct a numerical interface flux $\underline{\hat{f}}$ over a face with normal $\underline{n}$:
\begin{equation}\label{eq:fds-interface-flux}
\renewcommand{\arraystretch}{1.25}
    \underline{\hat{f}} = \{\underline{\tilde{f}}\}
    -\frac{1}{2}
    \Bigg[
    \mu_u|\tilde{U}|
    \begin{matrix} \begin{pmatrix}
        \Delta \tilde{\rho}  \\
        \Delta \tilde{\rho}\tilde{\underline{u}} \\
        \Delta \tilde{\rho}\tilde{E}
    \end{pmatrix} \end{matrix}
    +
    \delta \tilde{U}
    \begin{matrix} \begin{pmatrix}
        \tilde{\rho}   \\
        \tilde{\rho}\tilde{\underline{u}} \\
        \tilde{\rho}\tilde{H}
    \end{pmatrix} \end{matrix}
    +
    \delta \tilde{p}
    \begin{matrix} \begin{pmatrix}
        0 \\
        \underline{n} \\
        \tilde{U}
    \end{pmatrix} \end{matrix}
    \Bigg]
    = \{\underline{\tilde{f}}\} - \underline{\tilde{f}}^d
\end{equation}
\begin{equation} \label{eq:fds-interface-delta-up}
    \begin{split}
        \delta \tilde{U} & = \frac{\mu_{11}}{\tilde{\rho}|\tilde{v}|}\Delta\tilde{p} + \frac{\tilde{U}}{|\tilde{U}|}\mu_{12}\Delta\tilde{U} \\
        \delta \tilde{p} & = \frac{\tilde{U}}{|\tilde{U}|}\mu_{21}\Delta\tilde{p} + \tilde{\rho}|\tilde{v}|\mu_{22}\Delta\tilde{U} 
    \end{split}
\end{equation}
Where $\{(\dotp)\}=\frac{1}{2}((\dotp)_L+(\dotp)_R)$ and $\Delta(\dotp)=(\dotp)_R-(\dotp)_L$ are the central average and the jump between the left/right states of the interface and $U=\underline{u}\dotp\underline{n}$ is the interface-normal velocity.
The first term $\{\underline{\tilde{f}}\}$ is the central approximation to the physical flux, and the other terms are the artificial diffusion $\underline{\tilde{f}}^d$.
The first of these is the convective upwinding on the conserved variables.
The interface velocity perturbation $\delta \tilde{U}$ comes from the diffusion on the pressure in the entropy variables, and the interface pressure perturbation $\delta \tilde{p}$ comes from the diffusion on the velocity in the entropy variables.
The switch $\tilde{U}/|\tilde{U}|$ ensures that the off-diagonal terms maintain a diffusive character.
Matrix Rusanov fluxes such as the Roe scheme \cite{roe_approximate_1981} can be arranged in the form of (\ref{eq:fds-interface-flux}) \cite{liu_upwind_1989,weiss_preconditioning_1995} to compare existing Roe-type schemes at low Mach number to the scalings in table \ref{tab:diffusion-scaling}, as was done in \cite{li_mechanism_2013} and \cite{hope-collins_artificial_2022}.
This form will be referred to as the Roe form for the rest of the paper.

\section{Convection-pressure flux-vector splittings at low Mach number}\label{sec:cp-splittings}

The form (\ref{eq:fds-interface-flux},\ref{eq:fds-interface-delta-up}) is useful for analysing interface fluxes as it explicitly separates the non-diffusive central component from the stabilising artificial diffusion components, however the artificial diffusion in convection-pressure flux-vector splitting schemes is usually implicit in the formulation.
In this section, we construct approximate forms for three families of convection-pressure flux-vector splittings in the conserved variables that explicitly separate the artificial diffusion terms, so that we can make comparison to (\ref{eq:fds-interface-flux},\ref{eq:fds-interface-delta-up}).
We then transform the splittings to the entropy variables to make direct comparison to (\ref{eq:euler_modified}) and the results of \cite{hope-collins_artificial_2022}.

\subsection{Conserved variables form of convection-pressure flux-vector splittings}\label{sec:cp-splittings-conserved}

Convection-pressure flux-vector splittings split the physical flux vector $\underline{f}$ into two components:
\begin{equation}\label{eq:general-cp-flux-splitting}
    \underline{f} = U\underline{\varphi} + \underline{P}
\end{equation}
where $U$ is the local convecting velocity, $\underline{\varphi}$ is some vector of convected quantities, and $\underline{P}$ is some vector representing the pressure flux.
The choice of $\underline{\varphi}$ and $\underline{P}$ specify the flux-splitting.
For the three flux-splittings considered in this paper, the pressure flux can be written as $\underline{P}=p\underline{\Pi}$, where $p$ is the local pressure.
A numerical interface flux can be constructed from the flux-splitting (\ref{eq:general-cp-flux-splitting}) with the form:
\begin{equation}\label{eq:general-cp-interface-flux}
    \underline{\hat{f}} = U_{1/2}\underline{\varphi} + \underline{P}_{1/2}
\end{equation}
where $U_{1/2}$ is an interface velocity and $\underline{P}_{1/2}$ is an interface pressure flux.
We assume that the interface velocity and pressure can be rearranged (possibly approximately) into the form:
\begin{equation}\label{eq:general-interface-up}
    U_{1/2} = \{U\} - \delta U,
    \quad
    \underline{P}_{1/2} = \{\underline{P}\} - \delta\underline{P}
\end{equation}
where the interface velocity and pressure perturbations $\delta U$ and $\delta\underline{P}$ are diffusion terms similar to those in (\ref{eq:fds-interface-delta-up}).
The interface flux (\ref{eq:general-cp-interface-flux}) can then be rearranged into a form that separates the central and diffusive components by assuming $U_{1/2}>0$ and $\underline{\varphi}$ is taken as the upwind value based on $U_{1/2}$, as is usually the case.
Adding $(U_R\underline{\varphi}_R - U_R\underline{\varphi}_R)$ to the right hand side, we obtain:
\begin{equation}\label{eq:discrete-cp-diffusion}
    \underline{\hat{f}} = \{\underline{f}\} - \Big(\frac{U_R}{2}\Delta\underline{\varphi} + \delta U\underline{\varphi}_L + \delta\underline{P}\Big)
\end{equation}
The terms in brackets are all artificial diffusion terms.
Assuming smooth flow, the differences between left and right values are small, and we can approximate the flux as:
\begin{equation}\label{eq:general-cp-diffusion}
    \underline{\hat{f}} = \{\underline{f}\} - \Big(\frac{U}{2}\Delta\underline{\varphi} + \delta U\underline{\varphi} + \delta\underline{P}\Big)
\end{equation}
where $U$ and $\underline{\varphi}$ are some consistent approximation of the local values.
Comparing to equations (\ref{eq:fds-interface-flux},\ref{eq:fds-interface-delta-up}) in section \ref{sec:artificial-diffusion}, it can be seen that equation (\ref{eq:general-cp-diffusion}) closely approximates the Roe form of the artificial diffusion.
The first diffusion term is the convective upwinding, the second term is diffusion due to the interface velocity perturbation, and the third term is due to the interface pressure flux perturbation.
The interface velocity perturbation $\delta U$ is assumed to have the same form as in the Roe diffusion (\ref{eq:fds-interface-delta-up}), however the form of interface pressure flux perturbation $\delta\underline{P}$ can be different for each splitting.
The recasting of the convection term into central and upwinding components was shown for AUSM in the original paper from Liou \& Steffen \cite{liou_new_1993}, and the splitting of the pressure term was also used by Liou \cite{liou_sequel_1996} and Edwards \& Liou \cite{edwards_low-diffusion_1998,liou_ausm_1999,edwards_reflections_2019}.
The CUSP scheme \cite{jameson_artificial_1993} uses explicit diffusion terms with a similar form.
Now we have an explicit expression for the artificial diffusion we can substitute the specific forms of $\underline{\varphi}$ and $\underline{P}$ for each flux splitting.

\subsubsection{Liou-Steffen}
For the Liou-Steffen splitting we have:
\begin{equation}\label{eq:ls-splitting}
    \underline{\varphi} = 
    \begin{matrix}
    \begin{pmatrix}
        \rho \\ \rho u \\ \rho v \\ \rho H
    \end{pmatrix}
    \end{matrix},
    \quad
    \underline{P} = 
    p\underline{\Pi} =
    p
    \begin{matrix}
    \begin{pmatrix}
        0 \\ n_x \\ n_y \\ 0
    \end{pmatrix}
    \end{matrix},
    \quad
    \delta \underline{P} = \delta p \underline{\Pi}
\end{equation}
where $\delta p$ is the same interface pressure perturbation as the Roe diffusion (\ref{eq:fds-interface-delta-up}).
Substituting (\ref{eq:ls-splitting}) into (\ref{eq:general-cp-diffusion}), the approximate form of the diffusion is:
\begin{equation}\label{eq:ls-diffusion}
    \underline{\hat{f}} = \{\underline{f}\}
    - \frac{U}{2}
    \begin{matrix}
    \begin{pmatrix}
        \Delta\rho \\ \Delta\rho u \\ \Delta\rho v \\ \Delta\rho H
    \end{pmatrix}
    \end{matrix}
    - \delta U
    \begin{matrix}
    \begin{pmatrix}
        \rho \\ \rho u \\ \rho v \\ \rho H
    \end{pmatrix}
    \end{matrix}
    - \delta p
    \begin{matrix}
    \begin{pmatrix}
        0 \\ n_x \\ n_y \\ 0
    \end{pmatrix}
    \end{matrix}
\end{equation}
The first term, the convective upwinding, is almost exactly the same as the Roe diffusion, except that the energy equation is upwinded on jumps in the total enthalpy, not total energy.
For low Mach number flows, the zeroth order pressure is a factor of $\mathcalOM{\-2}$ larger than the kinetic energy, so to leading order the total energy and total enthalpy differ by a factor of $\gamma$.
The second term, the velocity perturbation diffusion, is exactly that of the Roe diffusion, which is particularly important for preventing pressure instabilities in purely convective flows.
The third term, the pressure perturbation diffusion, naturally has the form $\delta p \underline{\Pi}$ and matches the Roe form for the momentum equations but provides no diffusion on the energy equation, unlike the Roe form.

\subsubsection{Zha-Bilgen}
For the Zha-Bilgen splitting we have:
\begin{equation}\label{eq:zb-splitting}
    \underline{\varphi} = 
    \begin{matrix}
    \begin{pmatrix}
        \rho \\ \rho u \\ \rho v \\ \rho E
    \end{pmatrix}
    \end{matrix},
    \quad
    \underline{P} = 
    p\underline{\Pi} =
    p
    \begin{matrix}
    \begin{pmatrix}
        0 \\ n_x \\ n_y \\ U
    \end{pmatrix}
    \end{matrix},
    \quad
    \delta \underline{P} =
    \begin{matrix}
    \begin{pmatrix}
        0 \\ \delta p\,n_x \\ \delta p\,n_y \\ \delta(pU)
    \end{pmatrix}
    \end{matrix},
\end{equation}
leading to the approximate form:
\begin{equation}\label{eq:zb-diffusion1}
    \underline{\hat{f}} = \{\underline{f}\}
    - \frac{U}{2}
    \begin{matrix}
    \begin{pmatrix}
        \Delta\rho \\ \Delta\rho u \\ \Delta\rho v \\ \Delta\rho E
    \end{pmatrix}
    \end{matrix}
    - \delta U
    \begin{matrix}
    \begin{pmatrix}
        \rho \\ \rho u \\ \rho v \\ \rho E
    \end{pmatrix}
    \end{matrix}
    -
    \begin{matrix}
    \begin{pmatrix}
        0 \\ \delta p\,n_x \\ \delta p\,n_y \\ \delta(pU)
    \end{pmatrix}
    \end{matrix}
\end{equation}
The convective upwinding term is exactly that of the Roe form, so can be expected to give good resolution for the convective features.
The velocity perturbation term matches the Roe term in the mass and momentum equations, but uses the total energy scale in the energy equation instead of the total enthalpy.
The third term matches the Roe form except possibly in the energy equation, which depends on the specific expression for $\delta(pU)$.
If we use the expression:
\begin{equation}\label{eq:delta-pu-simplification}
    \delta(pU) = U\delta p
\end{equation}
then we can find a first form for the diffusion in terms of $\delta U$ and $\delta p$:
\begin{equation}\label{eq:zb-diffusion2}
    \underline{\hat{f}} = \{\underline{f}\}
    - \frac{U}{2}
    \begin{matrix}
    \begin{pmatrix}
        \Delta\rho \\ \Delta\rho u \\ \Delta\rho v \\ \Delta\rho E
    \end{pmatrix}
    \end{matrix}
    - \delta U
    \begin{matrix}
    \begin{pmatrix}
        \rho \\ \rho u \\ \rho v \\ \rho E
    \end{pmatrix}
    \end{matrix}
    - \delta p
    \begin{matrix}
    \begin{pmatrix}
        0 \\ n_x \\ n_y \\ U
    \end{pmatrix}
    \end{matrix}
\end{equation}
where the pressure perturbation term now has the form $\delta p \underline{\Pi}$, and is exactly that of the Roe scheme.
A second form is found if the $\delta(pU)$ term is instead written as:
\begin{equation}\label{eq:delta-pu-simplification-2}
    \delta(pU) = p\delta U + U\delta p
\end{equation}
Then the total diffusion on the energy equation is:
\begin{equation}\label{eq:zb-diffusion3}
\begin{split}
    \underline{\hat{f}}^d_{\rho E} & = U\Delta\rho E + (\rho E + p)\delta U + U\delta p \\
                                   & = U\Delta\rho E + \rho H \delta U + U\delta p
\end{split}
\end{equation}
which matches that in the Roe-type scheme (\ref{eq:fds-interface-flux}).

\subsubsection{Toro-Vasquez}
For the Toro-Vasquez splitting we have:
\begin{equation}\label{eq:tv-splitting}
\renewcommand{\arraystretch}{1.25}
    \underline{\varphi} = 
    \begin{matrix}
    \begin{pmatrix}
        \rho \\ \rho u \\ \rho v \\ \rho k
    \end{pmatrix}
    \end{matrix},
    \quad
    \underline{P} = 
    p\underline{\Pi} =
    p
    \begin{matrix}
    \begin{pmatrix}
        0 \\ n_x \\ n_y \\ \dfrac{\gamma U}{\gamma-1}
    \end{pmatrix}
    \end{matrix},
    \quad
    \delta \underline{P} =
    \begin{matrix}
    \begin{pmatrix}
        0 \\ \delta p\,n_x \\ \delta p\,n_y \\ \dfrac{\gamma\delta(pU)}{\gamma-1}
    \end{pmatrix}
    \end{matrix},
\end{equation}
leading to the approximate form:
\begin{equation}\label{eq:tv-diffusion1}
\renewcommand{\arraystretch}{1.25}
    \underline{\hat{f}} = \{\underline{f}\}
    - \frac{U}{2}
    \begin{matrix}
    \begin{pmatrix}
        \Delta\rho \\ \Delta\rho u \\ \Delta\rho v \\ \Delta\rho k
    \end{pmatrix}
    \end{matrix}
    - \delta U
    \begin{matrix}
    \begin{pmatrix}
        \rho \\ \rho u \\ \rho v \\ \rho k
    \end{pmatrix}
    \end{matrix}
    -
    \begin{matrix}
    \begin{pmatrix}
        0 \\ \delta p\,n_x \\ \delta p\,n_y \\ \dfrac{\gamma\delta(pU)}{\gamma-1}
    \end{pmatrix}
    \end{matrix}
\end{equation}
The Toro-Vasquez diffusion has the least similarity to the Roe form.
The diffusion on the continuity and momentum equations match, but the energy equation diffusion differs in all three terms.
As stated above, the ratio of the kinetic energy to the internal energy is $\mathcalOM{2}$, so the convective upwind diffusion and velocity perturbation diffusion on the energy equation in this form of the Toro-Vasquez splitting will be significantly smaller than in the Liou-Steffen and Zha-Bilgen splittings.
If we again use the first expression for $\delta(pU)$, equation (\ref{eq:delta-pu-simplification}), then the first form of the diffusion in terms of $\delta U$ and $\delta p$ is:
\begin{equation}\label{eq:tv-diffusion2}
\renewcommand{\arraystretch}{1.25}
    \underline{\hat{f}} = \{\underline{f}\}
    - \frac{U}{2}
    \begin{matrix}
    \begin{pmatrix}
        \Delta\rho \\ \Delta\rho u \\ \Delta\rho v \\ \Delta\rho k
    \end{pmatrix}
    \end{matrix}
    - \delta U
    \begin{matrix}
    \begin{pmatrix}
        \rho \\ \rho u \\ \rho v \\ \rho k
    \end{pmatrix}
    \end{matrix}
    - \delta p
    \begin{matrix}
    \begin{pmatrix}
        0 \\ n_x \\ n_y \\ \dfrac{\gamma U}{\gamma-1}
    \end{pmatrix}
    \end{matrix}
\end{equation}
The third diffusion term now has the form $\delta p\underline{\Pi}$, and matches the Roe form, up to a factor $\frac{\gamma}{\gamma-1}$ in the energy equation.
A second form of the Toro-Vasquez diffusion can be found similarly to the second Zha-Bilgen form by using (\ref{eq:delta-pu-simplification-2}).
This results in the total diffusion on the energy equation having the form:
\begin{equation}\label{eq:tv-diffusion3}
\begin{split}
    \underline{\hat{f}}^d_{\rho E} & = U\Delta\rho k + \Big(\rho k + \frac{\gamma p}{\gamma-1}\Big)\delta U + \frac{\gamma U\delta p}{\gamma-1} \\
                                   & = U\Delta\rho k + \rho H \delta U + \frac{\gamma U\delta p}{\gamma-1}
\end{split}
\end{equation}
where the velocity perturbation now matches the Roe form, but the convective upwinding and the pressure perturbation terms remain unchanged.\\

In this section it has been shown how convection-pressure flux-vector splittings can be approximately reformulated to make the artificial diffusion terms explicit, allowing comparison with the findings of the first paper.
This reformulation requires a number of assumptions which limit the scope of its applicability.
The first two assumptions are that the convection term in (\ref{eq:general-cp-interface-flux}) can be cast as $U_{1/2}\underline{\varphi}$, and that the interface velocity and pressure flux can be cast as the sum of a central approximation and a diffusive term (\ref{eq:general-interface-up}), which hold true for many schemes in the literature.
In this study we are interested in the asymptotic scaling of the artificial diffusion coefficients $\mu_{ij}$, and not their specific form.
This means that, for the purpose of our analysis, the reformulation into (\ref{eq:general-cp-interface-flux},\ref{eq:general-interface-up}) can be approximate and these first two assumptions are not particularly limiting even when they do not hold exactly.
The next assumptions are on the form of the pressure diffusion term $\delta\underline{P}$.
Firstly, that this term has the same sparsity as the pressure flux $\underline{P}$.
While this may seem like a logical assumption, we will see later that it does not apply to all schemes in the literature.
The second assumption is the simplification $\delta(pU)=U\delta p$ (\ref{eq:delta-pu-simplification}), which allows the pressure diffusion term of the Zha-Bilgen and Toro-Vasquez splittings to be written as $\delta\underline{P}=\delta p\underline{\Pi}$, as it naturally is for the Liou-Steffen splitting and the Roe flux.
However, (\ref{eq:delta-pu-simplification}) will not necessarily hold for all schemes even if the sparsity assumption holds, in which case the second forms of the Zha-Bilgen and Toro-Vasquez splittings may be more applicable.
Lastly, we note that the only assumption which is specific to low Mach number flow is that the flow is smooth, which we used in the step from (\ref{eq:discrete-cp-diffusion}) to (\ref{eq:general-cp-diffusion}).

The assumptions on the diffusion terms enforce only light requirements, and result in diffusion forms which closely match that of the Roe scheme so allow better comparison to the findings of \cite{hope-collins_artificial_2022}.
The approximate diffusion terms of all of these forms match the Roe form on the continuity and momentum equations, although they all differ to some extent on the energy equation.
The Liou-Steffen and both of the Zha-Bilgen forms match the Roe form of the artificial diffusion exactly in at least one of the terms (all three in the case of the second Zha-Bilgen form (\ref{eq:zb-diffusion3})), and approximately match in the others.
On the other hand, the Toro-Vasquez splitting shows little resemblance to the Roe form in the energy equation, other than in the second form (\ref{eq:tv-diffusion3}) where the velocity perturbation term matches.

\subsection{Entropy variables form}
The artificial diffusion fluxes (\ref{eq:ls-diffusion}), (\ref{eq:zb-diffusion2}), and (\ref{eq:tv-diffusion2}) are now transformed to the non-dimensional entropy variables, and the resulting modified equations are compared to (\ref{eq:euler_modified}) and the findings of the previous paper.
The Roe type diffusion (\ref{eq:fds-interface-flux},\ref{eq:fds-interface-delta-up}) in the entropy variables is:
\begin{equation}\label{eq:entropy-diffusion-fds}
    \mu_u|u|
    \begin{matrix} \begin{pmatrix}
        \phantom{\rho}\partial_{xx}p \\
        \rho\partial_{xx}u \\
        \rho\partial_{xx}v \\
        \phantom{\rho}\partial_{xx}s
        \end{pmatrix} \end{matrix}
    +
    \delta U
    \begin{matrix} \begin{pmatrix}
        \gamma p \\
        0 \\
        0 \\
        0
    \end{pmatrix} \end{matrix}
    +
    \delta p
    \begin{matrix} \begin{pmatrix}
        0 \\
        1 \\
        0 \\
        0
    \end{pmatrix} \end{matrix}
\end{equation}
which is exactly the right hand side of (\ref{eq:euler_modified}) written in terms of $\delta U$ and $\delta p$.
The diffusion terms found from the Liou-Steffen splitting (\ref{eq:ls-diffusion}) are:
\begin{equation}\label{eq:entropy-diffusion-ls}
    \mu_u|u|
    \begin{matrix} \begin{pmatrix}
        \gamma\partial_{xx}p \\
        \rho\partial_{xx}u \\
        \rho\partial_{xx}v \\
        \partial_{xx}s + (\gamma-1)\partial_{xx}p
    \end{pmatrix} \end{matrix}
    +
    \delta U
    \begin{matrix} \begin{pmatrix}
        \gamma p \\
        0 \\
        0 \\
        0
    \end{pmatrix} \end{matrix}
    +
    \delta p
    \begin{matrix} \begin{pmatrix}
        -M^2u(\gamma-1) \\
        1 \\
        0 \\
        -M^2u(\gamma-1)
    \end{pmatrix} \end{matrix}
\end{equation}
As expected, the second term in (\ref{eq:entropy-diffusion-ls}) exactly matches that of the Roe scheme.
The convective upwinding is slightly different to the Roe form - increased by a factor of $\gamma$ on the pressure, and the upwinding on the entropy field also has a contribution from the pressure variations.
Interestingly, the upwinding terms on the entropy field are equivalent to temperature diffusion: $ds + (\gamma-1)dp = \rho R\gamma dT$.
The pressure perturbation diffusion in the third term, which should act only on the velocity field, now also has anti-diffusion terms on the pressure and entropy fields, although these terms are reduced by a factor of $M^2$.

For the first form of the Zha-Bilgen splitting (\ref{eq:zb-diffusion2}) we have:
\begin{equation}\label{eq:entropy-diffusion-zb}
    \mu_u|u|
    \begin{matrix} \begin{pmatrix}
        \phantom{\rho}\partial_{xx}p \\
        \rho\partial_{xx}u \\
        \rho\partial_{xx}v \\
        \phantom{\rho}\partial_{xx}s
    \end{pmatrix} \end{matrix}
    +
    \delta U
    \begin{matrix} \begin{pmatrix}
        p \\
        0 \\
        0 \\
        -(\gamma-1)p
    \end{pmatrix} \end{matrix}
    +
    \delta p
    \begin{matrix} \begin{pmatrix}
        0 \\
        1 \\
        0 \\
        0
    \end{pmatrix} \end{matrix}
\end{equation}
As expected, the first and third terms in (\ref{eq:entropy-diffusion-zb}) exactly match those in the Roe scheme.
However, the $\delta U$ diffusion is a factor of $\gamma$ smaller on the pressure field, and has introduced an anti-diffusion term on the entropy field.
By inspection, the second Zha-Bilgen form (\ref{eq:zb-diffusion3}) will appear the same as the Roe form (\ref{eq:entropy-diffusion-fds}), with the correct pressure diffusion $\delta U$ term.\\
Finally, for the first form of the Toro-Vasquez splitting (\ref{eq:tv-diffusion2}) we have:
\begin{equation}\label{eq:entropy-diffusion-tv}
    \mu_u|u|
    \begin{matrix} \begin{pmatrix}
        0 \\
        \rho\partial_{xx}u \\
        \rho\partial_{xx}v \\
        \partial_{xx}s - \partial_{xx}p
    \end{pmatrix} \end{matrix}
    +
    \delta U
    \begin{matrix} \begin{pmatrix}
        0 \\
        0 \\
        0 \\
        -\gamma p
    \end{pmatrix} \end{matrix}
    +
    \delta p
    \begin{matrix} \begin{pmatrix}
        M^2u \\
        1 \\
        0 \\
        M^2u
    \end{pmatrix} \end{matrix}
\end{equation}
None of the three terms in (\ref{eq:entropy-diffusion-tv}) match those in (\ref{eq:entropy-diffusion-fds}).
The upwinding and $\delta U$ term provide no diffusion on the pressure field, which only has (erroneous) diffusion from the $\delta p$ term scaled by $M^2$.
The upwinding on the entropy field has a contribution from the pressure variations, similar to the Liou-Steffen form.
This time, the upwinding on the entropy field is equivalent to density diffusion: $ds - dp = -a^2d\rho$.\footnote{The negative sign does not indicate anti-diffusion because $\frac{\partial s}{\partial\rho}<0$.}
The only effect of the $\delta U$ term is an anti-diffusion term on the entropy equation, similar to that in first form of the Zha-Bilgen splitting.
By inspection, the second Toro-Vasquez form will appear as (\ref{eq:entropy-diffusion-tv}) but with the velocity perturbation form of the Roe scheme, which removes the pressure anti-diffusion term from the entropy equation and restores it to the pressure equation.\\

Because all forms are consistent with the Roe form on the mass and momentum equations in the conserved variables, they all have the correct diffusion on the velocity and vorticity equations in the entropy variables.
However, the diffusion on the pressure and entropy fields does not fully match the Roe form for any of the flux-splitting forms, apart from the second Zha-Bilgen form.

\subsection{Limit equations of flux-vector splittings}

In this section the limiting forms of the modified equations of each diffusion scheme are found for the convective and acoustic asymptotic regimes.
The limit equations are found with the following steps:
\begin{enumerate}
    \item Replace the right hand side of the modified equations (\ref{eq:euler_modified}) with one of the diffusion schemes (\ref{eq:entropy-diffusion-fds}-\ref{eq:entropy-diffusion-tv}).
    \item Set the scaling of $\mu_{11}$, $\mu_{22}$ according to either the convective or mixed diffusion scaling in table \ref{tab:diffusion-scaling}.
    \item Enforce the scaling of the physical quantities and their gradients according to either the convective or acoustic regime scaling in table \ref{tab:lowmach_scaling}.
    \item Retain only the leading order terms in each equation.
\end{enumerate}
The limit equations provide less detail than full asymptotic expansions of the modified equations would, but are significantly more compact and still give useful insight into the asymptotic behaviour of schemes at low Mach number, as shown in \cite{hope-collins_artificial_2022}.
The velocity equations for all three flux splittings match the Roe form which means that with the acoustic diffusion scaling they will all be inaccurate for convective features.
For this reason only the limit equations of convective and mixed diffusion scalings will considered.
To reduce the size of the relations, all schemes are taken to have asymptotically diagonal diffusion Jacobians i.e. $\mu_{12}\sim\mu_{21}\sim o(M^0)$, which is almost always the case for convection-pressure flux-vector splittings at low Mach number.

\subsubsection{Roe form}
The limit equations for the `ideal' modified equations (\ref{eq:euler_modified}) are included here for the flux-vector-splitting schemes to be compared against.
These limit equations are equivalent to those presented in section 3 of \cite{hope-collins_artificial_2022}, except with the specific form of the diffusion coefficients from the present equation (\ref{eq:euler_modified}).
Because (\ref{eq:euler_modified}) are the modified equations for Roe-type schemes in the entropy variables, we shall refer to them as the Roe form of the modified equations.
Their main features are described here; see \cite{hope-collins_artificial_2022} for more in-depth discussion.\\
The limit equations for the Roe form (\ref{eq:entropy-diffusion-fds}) using the convective diffusion scaling and enforcing convective variations from table \ref{tab:lowmach_scaling} are:
\begin{equation} \label{eq:limit-fds-Ac-convective}
    \begin{aligned}
        \partial_t\ord{p}{0} + \gamma\ord{p}{0}\partial_x\ord{u}{0} & = M^{\-2}\frac{\gamma\ord{p}{0}}{\ord{\rho}{0}|v|}\mu_{11}\partial_{xx}\ord{p}{2} \\
        \ord{\rho}{0}\partial_t\ord{u}{0} + \partial_x\ord{p}{2} + \ord{\rho u}{0}\partial_x\ord{u}{0} & = \ord{\rho}{0}\mu_u|u|\partial_{xx}\ord{u}{0} + \ord{\rho}{0}|v|\mu_{22}\partial_{xx}\ord{u}{0} \\
        \partial_t\ord{v}{0}                           +      \ord{u}{0}\partial_x\ord{v}{0} & = \mu_u|u|\partial_{xx}\ord{v}{0} \\
        \partial_t\ord{s}{0}                           +      \ord{u}{0}\partial_x\ord{s}{0} & = \mu_u|u|\partial_{xx}\ord{s}{0}
    \end{aligned}
\end{equation}
The limit equations for the Roe form (\ref{eq:entropy-diffusion-fds}) using the convective diffusion scaling and enforcing acoustic variations from table \ref{tab:lowmach_scaling} are:
\begin{equation} \label{eq:limit-fds-Ac-acoustic}
    \begin{aligned}
        0 & = M^{\-2}\frac{\gamma\ord{p}{0}}{\ord{\rho}{0}|v|}\mu_{11}\partial_{xx}\ord{p}{1} \\
        \ord{\rho}{0}\partial_{\tau}\ord{u}{0} + \partial_x\ord{p}{1} & = 0 \\
        \partial_{\tau}\ord{v}{0} & = 0 \\
        \partial_{\tau}\ord{s}{0} & = 0
    \end{aligned}
\end{equation}
The left-hand-side of the convective limit equations match those of the single scale convective asymptotic expansion (\ref{eq:convective_timescale}) by construction.
The diffusion on the velocity equation is well balanced, and the vorticity and entropy equations appear as scalar advection-diffusion equations with upwind diffusion.
Note that at steady state the divergence does not disappear, but is equal to the pressure diffusion term.
As mentioned previously, this term damps pressure oscillations from the solution, preventing chequerboard modes.
In the acoustic limit, this term overwhelms the physical terms in the pressure equation, causing it to become a Poisson-like relation for the acoustic pressure $\ord{p}{1}$ which quickly damps out any acoustic waves from the solution.

The limit equations for the Roe form (\ref{eq:entropy-diffusion-fds}) using the mixed diffusion scaling and enforcing convective variations are:
\begin{equation} \label{eq:limit-fds-Am-convective}
    \begin{aligned}
        \partial_t\ord{p}{0} + \gamma\ord{p}{0}\partial_x\ord{u}{0} & = 0 \\
        \ord{\rho}{0}\partial_t\ord{u}{0} + \partial_x\ord{p}{2} + \ord{\rho u}{0}\partial_x\ord{u}{0} & = \ord{\rho}{0}\mu_u|u|\partial_{xx}\ord{u}{0} + \ord{\rho}{0}|v|\mu_{22}\partial_{xx}\ord{u}{0} \\
        \partial_t\ord{v}{0}                           +      \ord{u}{0}\partial_x\ord{v}{0} & = \mu_u|u|\partial_{xx}\ord{v}{0} \\
        \partial_t\ord{s}{0}                           +      \ord{u}{0}\partial_x\ord{s}{0} & = \mu_u|u|\partial_{xx}\ord{s}{0}
    \end{aligned}
\end{equation}
The limit equations for the Roe form (\ref{eq:entropy-diffusion-fds}) using the mixed diffusion scaling and enforcing acoustic variations are:
\begin{equation} \label{eq:limit-fds-Am-acoustic}
    \begin{aligned}
        \partial_{\tau}\ord{p}{1} + \gamma\ord{p}{0}\partial_x\ord{u}{0} & = M^{\-2}\frac{\gamma\ord{p}{0}}{\ord{\rho}{0}|v|}\mu_{11}\partial_{xx}\ord{p}{1} \\
        \ord{\rho}{0}\partial_{\tau}\ord{u}{0} + \partial_x\ord{p}{1} & = 0 \\
        \partial_{\tau}\ord{v}{0} & = 0 \\
        \partial_{\tau}\ord{s}{0} & = 0
    \end{aligned}
\end{equation}
Because the only difference between the convective and mixed diffusion scalings is in the pressure diffusion term, the limit equations for velocity, vorticity, and entropy with the mixed diffusion scaling are identical to those with the convective diffusion scaling, with only the pressure equation differing.
At the convective limit the pressure diffusion disappears from the continuity equation.
This allows for divergence-free solutions, consistent with the incompressible limit but, as stated previously, makes the scheme susceptible to pressure chequerboard instabilities.
At the acoustic limit, the pressure diffusion no longer overwhelms the physical terms.
The left-hand-side of the pressure and velocity equations are the equations for low Mach number acoustics.
The diffusion on the velocity vanishes asymptotically, making the mixed scaling susceptible to velocity instabilities at the acoustic limit.
The leading order entropy does not vary on the acoustic timescale, consistent with the approximation of isentropic acoustics.

Because the diffusion on the velocity and vorticity equations match the Roe scheme for all three flux-vector splittings, these limit equations will match those of the Roe form.
In the following sections, the limit equations for the vorticity will be omitted, but the limit equations for the velocity will be retained so that the pressure-velocity subsystem can be understood as a whole.

\subsubsection{Liou-Steffen splitting}

The limit equations for the Liou-Steffen form (\ref{eq:entropy-diffusion-ls}) using the convective diffusion scaling and enforcing convective variations are:
\begin{equation} \label{eq:limit-ls-Ac-convective}
    \begin{aligned}
        \partial_t\ord{p}{0} + \gamma\ord{p}{0}\partial_x\ord{u}{0} & = M^{\-2}\frac{\gamma\ord{p}{0}}{\ord{\rho}{0}|v|}\mu_{11}\partial_{xx}\ord{p}{2} \\
        \ord{\rho}{0}\partial_t\ord{u}{0} + \partial_x\ord{p}{2} + \ord{\rho u}{0}\partial_x\ord{u}{0} & = \ord{\rho}{0}\mu_u|u|\partial_{xx}\ord{u}{0} + \ord{\rho}{0}|v|\mu_{22}\partial_{xx}\ord{u}{0} \\
        \partial_t\ord{s}{0}                           +      \ord{u}{0}\partial_x\ord{s}{0} & = \mu_u|u|\partial_{xx}\ord{s}{0}
    \end{aligned}
\end{equation}
The limit equations for the Liou-Steffen form (\ref{eq:entropy-diffusion-ls}) using the convective diffusion scaling and enforcing acoustic variations are:
\begin{equation} \label{eq:limit-ls-Ac-acoustic}
    \begin{aligned}
        0 & = M^{\-2}\frac{\gamma\ord{p}{0}}{\ord{\rho}{0}|v|}\mu_{11}\partial_{xx}\ord{p}{1} \\
        \ord{\rho}{0}\partial_{\tau}\ord{u}{0} + \partial_x\ord{p}{1} & = 0 \\
        \partial_{\tau}\ord{s}{0} & = 0
    \end{aligned}
\end{equation}
The limit equations for the Liou-Steffen form (\ref{eq:entropy-diffusion-ls}) using the mixed diffusion scaling and enforcing convective variations are:
\begin{equation} \label{eq:limit-ls-Am-convective}
    \begin{aligned}
        \partial_t\ord{p}{0} + \gamma\ord{p}{0}\partial_x\ord{u}{0} & = 0 \\
        \ord{\rho}{0}\partial_t\ord{u}{0} + \partial_x\ord{p}{2} + \ord{\rho u}{0}\partial_x\ord{u}{0} & = \ord{\rho}{0}\mu_u|u|\partial_{xx}\ord{u}{0} + \ord{\rho}{0}|v|\mu_{22}\partial_{xx}\ord{u}{0} \\
        \partial_t\ord{s}{0}                           +      \ord{u}{0}\partial_x\ord{s}{0} & = \mu_u|u|\partial_{xx}\ord{s}{0}
    \end{aligned}
\end{equation}
The limit equations for the Liou-Steffen form (\ref{eq:entropy-diffusion-ls}) using the mixed diffusion scaling and enforcing acoustic variations are:
\begin{equation} \label{eq:limit-ls-Am-acoustic}
    \begin{aligned}
        \partial_{\tau}\ord{p}{1} + \gamma\ord{p}{0}\partial_x\ord{u}{0} & = M^{\-2}\frac{\gamma\ord{p}{0}}{\ord{\rho}{0}|v|}\mu_{11}\partial_{xx}\ord{p}{1} \\
        \ord{\rho}{0}\partial_{\tau}\ord{u}{0} + \partial_x\ord{p}{1} & = 0 \\
        \partial_{\tau}\ord{s}{0} & = 0
    \end{aligned}
\end{equation}
The erroneous diffusion from the $\delta p$ terms vanishes asymptotically from the pressure and entropy equations for both the convective and mixed diffusion scalings at both the convective and acoustic limits.
The pressure diffusion in the entropy equation from the convective upwinding term also vanishes at both limits.
As a result, the limit equations for the Liou-Steffen form exactly match those of the Roe form.
This should not be surprising given the success of low Mach number schemes based on this splitting with both convective and mixed diffusion scalings.

\subsubsection{Zha-Bilgen splitting}

The limit equations for the first Zha-Bilgen form (\ref{eq:entropy-diffusion-zb}) using the convective diffusion scaling and enforcing convective variations are:
\begin{equation} \label{eq:limit-zb-Ac-convective}
    \begin{aligned}
        \partial_t\ord{p}{0} + \gamma\ord{p}{0}\partial_x\ord{u}{0} & = M^{\-2}\frac{\ord{p}{0}}{\ord{\rho}{0}|v|}\mu_{11}\partial_{xx}\ord{p}{2} \\
        \ord{\rho}{0}\partial_t\ord{u}{0} + \partial_x\ord{p}{2} + \ord{\rho u}{0}\partial_x\ord{u}{0} & = \ord{\rho}{0}\mu_u|u|\partial_{xx}\ord{u}{0} + \ord{\rho}{0}|v|\mu_{22}\partial_{xx}\ord{u}{0} \\
        \partial_t\ord{s}{0}                           +      \ord{u}{0}\partial_x\ord{s}{0} & = \mu_u|u|\partial_{xx}\ord{s}{0} - M^{\-2}(\gamma-1)\frac{\ord{p}{0}}{\ord{\rho}{0}|v|}\mu_{11}\partial_{xx}\ord{p}{2} \\
    \end{aligned}
\end{equation}
The limit equations for the Zha-Bilgen form (\ref{eq:entropy-diffusion-zb}) using the convective diffusion scaling and enforcing acoustic variations are:
\begin{equation} \label{eq:limit-zb-Ac-acoustic}
    \begin{aligned}
    0 & = M^{\-2}\frac{\ord{p}{0}}{\ord{\rho}{0}|v|}\mu_{11}\partial_{xx}\ord{p}{1} \\
        \ord{\rho}{0}\partial_{\tau}\ord{u}{0} + \partial_x\ord{p}{1} & = 0 \\
        \partial_{\tau}\ord{s}{0} & = -M^{\-2}(\gamma-1)\frac{\ord{p}{0}}{\ord{\rho}{0}|v|}\mu_{11}\partial_{xx}\ord{p}{1}
    \end{aligned}
\end{equation}
The pressure diffusion is a factor of $\gamma$ smaller than for the Roe and Liou-Steffen forms, although if $\gamma\sim\mathcal{O}(1)$ then this should not have a significant effect, and can easily be compensated for in the definition of $\mu_{11}$.
Other than this difference, the pressure, velocity, and vorticity equations match the Roe and Liou-Steffen forms, so can be expected to perform similarly.
On the other hand, the entropy equation retains the pressure anti-diffusion term in both limits, which could cause erroneous entropy generation from leading order pressure variations.\\
The limit equations for the Zha-Bilgen form (\ref{eq:entropy-diffusion-zb}) using the mixed diffusion scaling and enforcing convective variations are:
\begin{equation} \label{eq:limit-zb-Am-convective}
    \begin{aligned}
        \partial_t\ord{p}{0} + \gamma\ord{p}{0}\partial_x\ord{u}{0} & = 0 \\
        \ord{\rho}{0}\partial_t\ord{u}{0} + \partial_x\ord{p}{2} + \ord{\rho u}{0}\partial_x\ord{u}{0} & = \ord{\rho}{0}\mu_u|u|\partial_{xx}\ord{u}{0} + \ord{\rho}{0}|v|\mu_{22}\partial_{xx}\ord{u}{0} \\
        \partial_t\ord{s}{0}                           +      \ord{u}{0}\partial_x\ord{s}{0} & = \mu_u|u|\partial_{xx}\ord{s}{0} \\
    \end{aligned}
\end{equation}
The limit equations for the Zha-Bilgen form (\ref{eq:entropy-diffusion-zb}) using the mixed diffusion scaling and enforcing acoustic variations are:
\begin{equation} \label{eq:limit-zb-Am-acoustic}
    \begin{aligned}
        \partial_{\tau}\ord{p}{1} + \gamma\ord{p}{0}\partial_x\ord{u}{0} & = M^{\-2}\frac{\ord{p}{0}}{\ord{\rho}{0}|v|}\mu_{11}\partial_{xx}\ord{p}{1} \\
        \ord{\rho}{0}\partial_{\tau}\ord{u}{0} + \partial_x\ord{p}{1} & = 0 \\
        \partial_{\tau}\ord{s}{0} & = 0
    \end{aligned}
\end{equation}
With the mixed diffusion scaling the pressure anti-diffusion term in the entropy equation vanishes asymptotically at both limits, so the Zha-Bilgen form matches the Roe form at the convective limit, and the only difference at the acoustic limit is the factor of $\gamma$ in the pressure equation diffusion.

The limit equations for the second Zha-Bilgen form (\ref{eq:zb-diffusion3}) will be identical to the limit equations for the Roe form, which will remove the erroneous pressure anti-diffusion term from the entropy equation at the convective limit.

\subsubsection{Toro-Vasquez splitting}

The limit equations for the first Toro-Vasquez form (\ref{eq:entropy-diffusion-tv}) using the convective diffusion scaling and enforcing convective variations are:
\begin{equation} \label{eq:limit-tv-Ac-convective}
    \begin{aligned}
        \partial_t\ord{p}{0} + \gamma\ord{p}{0}\partial_x\ord{u}{0} & = 0 \\
        \ord{\rho}{0}\partial_t\ord{u}{0} + \partial_x\ord{p}{2} + \ord{\rho u}{0}\partial_x\ord{u}{0} & = \ord{\rho}{0}\mu_u|u|\partial_{xx}\ord{u}{0} + \ord{\rho}{0}|v|\mu_{22}\partial_{xx}\ord{u}{0} \\
        \partial_t\ord{s}{0}                           +      \ord{u}{0}\partial_x\ord{s}{0} & = \mu_u|u|\partial_{xx}\ord{s}{0} - M^{\-2}\frac{\gamma\ord{p}{0}}{\ord{\rho}{0}|v|}\mu_{11}\partial_{xx}\ord{p}{2}
    \end{aligned}
\end{equation}
The limit equations for the Toro-Vasquez form (\ref{eq:entropy-diffusion-tv}) using the convective diffusion scaling and enforcing acoustic variations are:
\begin{equation} \label{eq:limit-tv-Ac-acoustic}
    \begin{aligned}
        \partial_{\tau}\ord{p}{1} + \gamma\ord{p}{0}\partial_x\ord{u}{0} & = 0 \\
        \ord{\rho}{0}\partial_{\tau}\ord{u}{0} + \partial_x\ord{p}{1} & = 0 \\
        \partial_{\tau}\ord{s}{0} & =  -M^{\-2}\frac{\gamma\ord{p}{0}}{\ord{\rho}{0}|v|}\mu_{11}\partial_{xx}\ord{p}{1}
    \end{aligned}
\end{equation}
The lack of pressure diffusion on the pressure equation means that, even with the convective diffusion scaling, the divergence relation is undamped and therefore susceptible to pressure chequerboard instabilities.
At the acoustic limit, the absence of this pressure diffusion means that the pressure equation does not become a Poisson-like relation for the acoustic pressure $\ord{p}{1}$ which would damp out any acoustic waves.
The erroneous $\delta p$ terms in the pressure and entropy equations vanish asymptotically at both limits, as does the pressure diffusion in the entropy equation from the convective upwinding term, leaving the correct entropy upwinding.
However, the pressure anti-diffusion term from $\delta U$ remains on the entropy equation, as for the Zha-Bilgen form.\\

The limit equations for the Toro-Vasquez form (\ref{eq:entropy-diffusion-tv}) using the mixed diffusion scaling and enforcing convective variations are:
\begin{equation} \label{eq:limit-tv-Am-convective}
    \begin{aligned}
        \partial_t\ord{p}{0} + \gamma\ord{p}{0}\partial_x\ord{u}{0} & = 0 \\
        \ord{\rho}{0}\partial_t\ord{u}{0} + \partial_x\ord{p}{2} + \ord{\rho u}{0}\partial_x\ord{u}{0} & = \ord{\rho}{0}\mu_u|u|\partial_{xx}\ord{u}{0} + \ord{\rho}{0}|v|\mu_{22}\partial_{xx}\ord{u}{0} \\
        \partial_t\ord{s}{0}                           +      \ord{u}{0}\partial_x\ord{s}{0} & = \mu_u|u|\partial_{xx}\ord{s}{0}
    \end{aligned}
\end{equation}
The limit equations for the Toro-Vasquez form (\ref{eq:entropy-diffusion-tv}) using the mixed diffusion scaling and enforcing acoustic variations are:
\begin{equation} \label{eq:limit-tv-Am-acoustic}
    \begin{aligned}
        \partial_{\tau}\ord{p}{1} + \gamma\ord{p}{0}\partial_x\ord{u}{0} & = 0 \\
        \ord{\rho}{0}\partial_{\tau}\ord{u}{0} + \partial_x\ord{p}{1} & = 0 \\
        \partial_{\tau}\ord{s}{0} & = 0
    \end{aligned}
\end{equation}
With the mixed diffusion scaling, the limit equations at the convective limit match the Roe form, however at the acoustic limit all diffusion terms vanish asymptotically, which leaves the scheme completely undamped.
It should be noted that the pressure diffusion in the pressure equation of the Roe form (\ref{eq:entropy-diffusion-fds}) with the mixed diffusion scaling still stabilises the zeroth and first order pressures $\ord{p}{0,1}$ even if it disappears from the limit equations at the convective limit (\ref{eq:limit-fds-Am-convective}).
This can be seen from an asymptotic expansion of the modified or discrete equations, even if it is not evident from the limit equations \cite{hope-collins_artificial_2022}.
The pressure diffusion in the pressure equation is completely absent from the first Toro-Vasquez form, which means that even though the limit equations (\ref{eq:limit-tv-Am-convective}) match those of the Roe scheme, it is potentially susceptible to instabilities which do not occur for the Roe scheme.

The limit equations for the second Toro-Vasquez form (\ref{eq:tv-diffusion3}) are, like those for the second Zha-Bilgen form, identical to the limit equations for the Roe form.
This not only removes the erroneous pressure anti-diffusion term from the entropy limit equations, it also restores the pressure diffusion terms to the pressure limit equations, which will stabilise the convective diffusion scaling at the convective limit, the mixed diffusion scaling at the acoustic limit, and will provide the proper damping on the acoustic variations for the convective diffusion scaling at the acoustic limit.

One last point to mention is that that erroneous $\delta p$ terms in pressure and entropy equations of the Liou-Steffen and Toro-Vasquez forms (\ref{eq:entropy-diffusion-ls}) and (\ref{eq:entropy-diffusion-tv}) vanish by a factor of $M^2$ from the limit equations, so these terms are unlikely to have any visible effect on the results found with these schemes.
On the other hand, the erroneous $\delta U$ terms in the entropy equations of the first Zha-Bilgen and Toro-Vasquez forms (\ref{eq:entropy-diffusion-zb}) and (\ref{eq:entropy-diffusion-tv}) only vanish by a factor of $M$ from the limit equations for the mixed diffusion scaling (\ref{eq:limit-zb-Am-convective},\ref{eq:limit-zb-Am-acoustic}) and (\ref{eq:limit-tv-Am-convective},\ref{eq:limit-tv-Am-acoustic}).
For moderately low Mach number, the pressure anti-diffusion terms in the entropy equations of the first Zha-Bilgen and Toro-Vasquez forms with mixed diffusion scaling may leave small but noticeable residues.\\

In this section we have shown that transforming to the entropy variables and examining the limiting forms of the modified equations provides additional insights into the structure of flux-vector-splittings at low Mach number.
The limit equations for the Liou-Steffen splitting exactly match those of the Roe form, despite only matching one of three diffusion terms exactly in the conserved variables.
The Zha-Bilgen splitting matches the Roe form on the pressure, velocity, and vorticity equations, up to a factor of $\gamma$ on the pressure diffusion.
However, with the convective diffusion scaling, a pressure anti-diffusion term remains in the entropy equations, which may lead to spurious entropy generation.
The Toro-Vasquez splitting has no diffusion on the pressure equation, which may lead to instabilities for both the convective and acoustic limits for any diffusion scaling.
Like for the Zha-Bilgen splitting, the Toro-Vasquez splitting with the convective diffusion scaling has a pressure anti-diffusion term on the entropy equation at both the convective and acoustic limits.
The second forms of the Zha-Bilgen and Toro-Vasquez splittings match the limit equations of the Roe scheme.

\section{Comparison of existing schemes}\label{sec:existing-schemes}

\begin{table}
    \centering
    \begin{tabularx}{\textwidth}{|p{0.25\textwidth}|c|c|c|X|} \hline
         \multirow{2}{*}{Scheme} & \multirow{2}{*}{Splitting} & \multicolumn{2}{c|}{Scaling}                                 & \multirow{2}{*}{Comments} \\ \cline{3-4}
                                 &                            & $\sigma_u\sim\mathcal{O}(1)$  & $\sigma_a\sim\mathcal{O}(1)$ &                           \\ \hline
         Zha-Bilgen 1993 \cite{zha_numerical_1993} & Zha-Bilgen & \multicolumn{2}{c|}{Acoustic}                      & Original Zha-Bilgen transonic scheme. \\ \hline
         AUSM/AUSM$^+$ 1993/1996 \cite{liou_new_1993,liou_sequel_1996} & Liou-Steffen & \multicolumn{2}{c|}{Acoustic}   & Original Liou-Steffen transonic scheme. \\ \hline
         LDFSS 1998/1999/2001 \cite{edwards_low-diffusion_1998,liou_numerical_1999,edwards_towards_2001} & Liou-Steffen & \multicolumn{2}{c|}{Convective} & AUSM with elements of flux-preconditioning. \\ \hline
         AUSM$^+$-up 2006 \cite{liou_sequel_2006} & Liou-Steffen & \multicolumn{2}{c|}{Convective}                   & Discrete asymptotic expansion to find scaling of diffusion coefficients. \\ \hline
         SLAU 2009/11 \cite{shima_parameter-free_2011} & Liou-Steffen & \multicolumn{2}{c|}{Mixed}                   & Numerically shows acoustic capability. \\ \hline
         Toro-Vasquez 2012 \cite{toro_flux_2012} & Toro-Vasquez & \multicolumn{2}{c|}{Acoustic}                      & Original Toro-Vasquez transonic scheme. \\ \hline
         Li \& Gu 2013 \cite{li_mechanism_2013} & Liou-Steffen & \multicolumn{2}{c|}{Mixed}                          & Modified AUSM$^+$-up to mixed scheme according to their guidelines. \\ \hline
         Sachdev et al. 2012 \cite{sachdev_improved_2012} & Liou-Steffen & Convective                    & Mixed     & Adaptive schemes for AUSM$^+$-up and SLAU. \\ \hline
         SLAU-WS 2013 \cite{shima_improvement_2013} & Liou-Steffen & \multicolumn{2}{c|}{Mixed/convective}                   & Modifies the pressure diffusion in SLAU to increase by a factor of $M$ in the presence of pressure `wiggles' independently of the timestep. \\ \hline
         TV-MAS 2017 \cite{sun_robust_2017} & Toro-Vasquez & \multicolumn{2}{c|}{Mixed}        & $U_{1/2}$ only adds upwind diffusion. $\underline{P}_{1/2}$ adds Zha-Bilgen $\delta U$ diffusion to the density and energy equations only, and adds Toro-Vasquez $\delta p$ diffusion to the momentum equations with mixed scaling and to the energy equation with the acoustic scaling. \\ \hline
         TV-MAS2 2018 \cite{lin_density_2018} & Toro-Vasquez & \multicolumn{2}{c|}{Mixed}        & Reduces $\delta p$ term in energy equation of TVMAS to the mixed scaling. \\ \hline
         E-AUSMPWAS 2018 \cite{qu_new_2018} & Zha-Bilgen & \multicolumn{2}{c|}{Mixed}                                & Uses $\delta(pU)=p\delta U$, resulting in second Zha-Bilgen form of $\delta U$ and Liou-Steffen form of $\delta p$. \\ \hline
         AUPM 2018 \cite{chen_novel_2018} & Zha-Bilgen & \multicolumn{2}{c|}{Mixed}                                  & Second Zha-Bilgen form. \\ \hline
         ZB-FVS-Corr 2020 \cite{iampietro_low-diffusion_2020} & Zha-Bilgen & \multicolumn{2}{c|}{Mixed}              & Second Zha-Bilgen form. \\ \hline
         AUSM-M 2020 \cite{chen_improved_2020} & Liou-Steffen & \multicolumn{2}{c|}{Mixed}                           & Reduces $\mu_{22}$ diffusion through shear layer. \\ \hline
         TVAP 2021 \cite{chen_low-diffusion_2021} & Toro-Vasquez & \multicolumn{2}{c|}{Mixed}                        & $U_{1/2}$ only adds upwind diffusion. $\underline{P}_{1/2}$ adds Liou-Steffen $\delta U$ diffusion to the density and energy equations only, and adds Liou-Steffen $\delta p$ diffusion to momentum equations.  \\ \hline
    \end{tabularx}
    \caption{Diffusion scaling of a number of existing low Mach number convection-pressure flux-vector splitting schemes. Some schemes have different scalings when the timestep is calculated using the convective CFL ($\sigma_u$) and $M_u\sim\mathcal{O}(M)$, or using the acoustic CFL ($\sigma_a$) and $M_u\sim\mathcal{O}(1)$.}
    \label{tab:existing-schemes}
\end{table}

In this section we compare the current analysis to previous studies and schemes, and identify the diffusion scaling of a number of existing schemes from the literature, which are shown in table \ref{tab:existing-schemes} along with the diffusion scaling they use.
Almost all of the studies on convection-pressure flux-vector splittings at low Mach number have focused on the Liou-Steffen splitting, which reflects its popularity over the last 30 years, and all of these studies were carried out in the conserved variables.\footnote{Liou 2017 \cite{liou_root_2017} swapped the energy equation for the entropy equation in a number of schemes including AUSM in the context of the overheating problem in high speed receding flow.}\\

Liou \& Edwards \cite{edwards_low-diffusion_1998,liou_ausm_1999,liou_numerical_1999,edwards_towards_2001} discuss the use of ideas from low Mach number flux preconditioning for improving the performance of Liou-Steffen splitting schemes at low Mach number, showing that the diffusion coefficients should become independent of the Mach number as the incompressible regime is approached.
This is equivalent to the convective diffusion scaling in table \ref{tab:diffusion-scaling}, which flux-difference splitting schemes using flux preconditioning also approach \cite{turkel_preconditioning_1999,guillard_behaviour_1999}.
An overview of these developments, and of the AUSM family of schemes more generally, is given by Edwards in \cite{edwards_towards_2001,edwards_reflections_2019}.
In the development of AUSM$^+$-up for all speeds \cite{liou_sequel_2006} Liou goes into detail - including discrete asymptotic analysis - on the design of an AUSM scheme for low Mach number independently from the preconditioned system (although inspired by the earlier work mentioned), and also arrives at the convective diffusion scaling.

Dellacherie 2010 and Dellacherie et al 2016 \cite{dellacherie_analysis_2010,dellacherie_construction_2016} analysed the requirements on Godunov schemes to accurately resolve convective flow features at low Mach number, showing that the reduction from $\mu_{22}\sim\mathcalOM{\-1}$ for the acoustic diffusion scaling to $\mu_{22}\sim\mathcalOM{0}$ for the mixed or convective diffusion scalings is essential.
It is also shown that AUSM/AUSM$^+$ \cite{liou_new_1993,liou_sequel_1996} are unsuitable for low Mach number flow, having the acoustic diffusion scaling on the pressure perturbation term, whereas AUSM$^+$-up has the correct scaling on $\mu_{22}$ so is suitable for low Mach number flow as designed.
By studying existing low Mach number Roe-type schemes, Li \& Gu 2013 \cite{li_mechanism_2013} draw out a set of guidelines for the design of such schemes.
By applying these guidelines to the AUSM$^+$-up scheme, they modified the velocity perturbation to change the diffusion scaling from convective to mixed.
More detail on how the approach used in the current paper and \cite{hope-collins_artificial_2022} relates to the works of Dellacherie and Li \& Gu can be found in \cite{hope-collins_artificial_2022}.

The works of Venkateswaran and co-workers \cite{venkateswaran_efficiency_2000,venkateswaran_artficial_2003,potsdam_unsteady_2007,sachdev_improved_2012} were, to the authors' knowledge, the earliest examples of the mixed diffusion scaling in the literature, and some of the only studies to explicitly identify the mixed diffusion scaling as a combination of the convective and acoustic diffusion scalings which is suitable for both limits.
In one of their more recent studies, Sachdev et al. \cite{sachdev_improved_2012} show the two instabilities that the mixed diffusion scaling is susceptible to: pressure chequerboards at the convective limit and a velocity grid mode at the acoustic limit.
They show that AUSM$^+$-up and SLAU \cite{shima_parameter-free_2011} have the convective and mixed diffusion scalings respectively, with the expected behaviour, and go on to show how they can both be modified into adaptive schemes i.e. approaching the convective diffusion scaling as $\Delta t\to\infty$ and approaching the mixed diffusion scaling as $\Delta t\to0$.
This adaptive behaviour is achieved using an ``unsteady Mach number'' $M_{u}=L_{\infty}/(\Delta t a_{\infty})$.
Shima 2013 \cite{shima_improvement_2013} takes a different approach to resolving this problem, damping out pressure instabilities by switching the diffusion scaling near pressure ``wiggles'' i.e. locations where the ratio of fourth to second derivatives of the pressure field is large.\\

From table \ref{tab:existing-schemes} it is clear that the Liou-Steffen splitting is the most commonly used, although the Zha-Bilgen and Toro-Vasquez splittings have recently gained more attention for low Mach number flow.
Just as for Roe-type schemes, the early Liou-Steffen schemes were rooted in the low Mach number preconditioning to obtain the correct scaling, which results in the convective diffusion scaling.
The almost total switch to the mixed diffusion scaling also happened around the same time as for Roe-type schemes \textit{c.}2008 - Postdam et al. 2007 \cite{potsdam_unsteady_2007}, Thornber et al. 2008 \cite{thornber_improved_2008,thornber_numerical_2008} and Li \& Gu 2008 \cite{li_all-speed_2008} for Roe schemes, and Shima \& Kitamura 2009/11 (SLAU) \cite{shima_new_2009,shima_parameter-free_2011} for Liou-Steffen schemes.
This means that, to the authors' knowledge, no Zha-Bilgen or Toro-Vasquez schemes using the convective diffusion scaling exist in the literature.
Of these earlier Roe-type fluxes with the mixed diffusion scaling, only Potsdam et al. had the explicit aim of creating a scheme capable of resolving both convective and acoustic low Mach number features, whereas aeroacoustic simulations were one of the original target applications for SLAU.\\


The AUSM$^+$-up scheme is one of the most popular convection-pressure flux-vector splitting schemes for low Mach number.
As stated earlier the diffusion coefficients become independent of the Mach number as $M\to0$, which means that it has the convective diffusion scaling.
As such, it can accurately resolve convective flow features in low Mach number flow \cite{liou_sequel_2006,dellacherie_analysis_2010} and is robust against the pressure chequerboard instabilities that can occur for collocated schemes for low Mach number or incompressible flow.
On the other hand, it damps acoustic variations very quickly \cite{moguen_pressurevelocity_2012,moguen_pressurevelocity_2013}, and suffers from a spectral radius which scales as $\mathcalOM{\-2}$ instead of the $\mathcalOM{\-1}$ scaling of the mixed and acoustic diffusion scalings and the Jacobian of the physical flux vector \cite{hope-collins_artificial_2022}.
The $\mathcalOM{\-2}$ spectral radius scaling of the convective diffusion scaling has long been known in the context of Roe-type schemes since it was first proven by Birken \& Meister 2005 \cite{birken_stability_2005}.
This spectral radius means that the stability bound on the timestep for an explicit scheme reduces prohibitively as $\Delta t\sim\mathcalOM{2}$, a factor of $M$ faster than usual.

Birken \& Meister noted that other schemes with the convective diffusion scaling would also suffer from these problems, and gave the specific example of the AUSMDV scheme \cite{wada_flux_1994}.
That these findings also apply to AUSM$^+$-up appears to have gone unrecognised (or at least unmentioned) in the more recent literature, although evidence of it can be seen in a number of studies.
In the original paper \cite{liou_sequel_2006} it can be seen from figure 22 that the convergence of the implicit scheme stalls after at least four orders of magnitude decrease in the residual, whereas convergence to machine zero is achieved when low Mach number preconditioning is used.
This is consistent with the expectation that the spectral radius and, more importantly, the condition number of the system is reduced from $\mathcalOM{\-2}$ to $\mathcalOM{0}$ by the preconditioning.
Several papers have reported stability issues with AUSM$^+$-up at very low Mach number when using an explicit timestepping strategy \cite{matsuyama_performance_2014,kitamura_reduced_2016,chen_improved_2018}.
The response to these issues has been to prevent the Mach number used in the calculation of the artificial diffusion falling below some cut-off Mach number $M_{co}>M_{\infty}$.
Keeping $M_{co}$ above the actual Mach number of the flow has two effects.
Firstly, it reduces the $\mu_{11}$ pressure diffusion, which reduces the severe scaling of the spectral radius and alleviates the stability restrictions as intended.
Secondly, it increases the $\mu_{22}$ velocity diffusion, which results in a more diffusive solution approaching that found with the acoustic diffusion scaling.
Li \& Gu \cite{li_mechanism_2013} mention that the AUSM$^+$-up scheme suffers from the cut-off Mach number problem just as the preconditioned Roe scheme does, but do not explicitly connect this back to the spectral radius scaling and the results of Birken \& Meister.
More recently, Chen et al. 2018 \cite{chen_improved_2018} showed empirically that the scaling of the maximum stable timestep of AUSM$^+$-up is what would be expected of the convective diffusion scaling,\footnote{This can be seen from table 1 and figure 5 in \cite{chen_improved_2018} which show linear scaling of the maximum timestep with the Mach number. This scaling is found by holding the speed of sound constant and varying the convective velocity to control the Mach number, so $\Delta t\sim\mathcalOM{}$ is equivalent to $\Delta t\sim\mathcalOM{2}$ when holding the convective velocity constant and varying the speed of sound.

Incidentally, the authors also empirically found this timestep scaling for AUSM$^+$-up, independently of \cite{chen_improved_2018}. Investigating if and how it was connected to the results of Birken \& Meister was one of the original motivations for \cite{hope-collins_artificial_2022} and the current study.} but again do not remark on the connection to the preconditioned Roe scheme or the results of Birken \& Meister.

The remedies to the stability issues of the convective scheme which do not sacrifice accuracy are the same as for the Roe-type schemes.
The most common solution is to use low Mach number preconditioning, which results in a $\mathcalOM{0}$ spectral radius on the convective timescale, but destroys the time accuracy of the scheme \cite{turkel_preconditioning_1999}.
When time accuracy is required, an implicit scheme can be used, often in conjunction with dual-time stepping \cite{jameson_time_1991}, which allows the use of low Mach number preconditioning while maintaining time accuracy \cite{venkateswaran_dual_1995,shima_cfd_2011,shima_new_2013}.
Using a scheme with the mixed diffusion scaling would also avoid this issue, but at the price of losing the robustness against pressure chequerboard instabilities.
It should be mentioned that preconditioning does not entirely remove the need for some cut-off Mach number because, although the condition number is reduced which improves the linear long-time convergence, the eigenvectors of the preconditioned system become increasingly non-normal as $M\to0$ which impedes the non-linear short-time convergence \cite{darmofal_importance_1996}.
In this case the cut-off Mach number should be applied only to the preconditioning matrix and $\mu_{11}$ to keep the correct scaling of $\mu_{22}$.\\

Turning now to the Zha-Bilgen and Toro-Vasquez splittings, the first thing to note is that none of the schemes in table \ref{tab:existing-schemes} use the first forms where $\delta(pU)=U\delta p$, which gives $\delta\underline{P}=\delta p\underline{\Pi}$ and maintains the separation of the $\delta p$ diffusion in the pressure perturbation term and the $\delta U$ diffusion in the velocity perturbation term.
Chen et al. 2018 (AUPM) \cite{chen_novel_2018} and Iampietro et al. 2020 \cite{iampietro_low-diffusion_2020} both use the second form of the Zha-Bilgen splitting with $\delta(pU)=p\delta U + U\delta p$, which gives diffusion equivalent to the Roe scheme on all three diffusion terms - convective upwinding, and the velocity and pressure perturbations.
Qu et al. 2018 (E-AUSMPWAS) \cite{qu_new_2018} use $\delta(pU)=p\delta U$, which is equivalent to using the Liou-Steffen form for both the velocity and pressure perturbation diffusion terms.
According to the findings of section \ref{sec:cp-splittings}, the differences between different forms of the $\delta p$ pressure perturbation terms are much less important than the form of the velocity perturbation term, so the difference in behaviour between these two expressions for $\delta(pU)$ should be small.

In the three Toro-Vasquez schemes in table \ref{tab:existing-schemes}, Sun et al. 2017 (TV-MAS) \cite{sun_robust_2017}, Lin et al. 2018 (TV-MAS2) \cite{lin_density_2018}, and Chen et al. 2021 (TVAP) \cite{chen_low-diffusion_2021}, the convection term $U_{1/2}\underline{\varphi}$ adds only convective upwind diffusion, and not the velocity perturbation $\delta U$ diffusion.
Instead, all of the $\delta U$ and $\delta p$ diffusion terms are included through the pressure perturbation term $\delta\underline{P}$.
$\delta U$ diffusion terms are added only to the density and energy equations, with the Liou-Steffen form used in the energy equation.
TVAP modifies this slightly, with the static enthalpy $(\rho h \delta U)$ used instead of the total enthalpy $(\rho E \delta U)$.
None of the TV schemes apply $\delta U$ diffusion to the momentum equations.
In their review of low Mach Roe-type schemes, Li \& Gu 2013 \cite{li_mechanism_2013} note that inconsistent application of the velocity perturbation diffusion across the different equations reduces the effectiveness of the convective diffusion scaling at suppressing pressure chequerboards.
Whether the same is true for the Toro-Vasquez flux with the mixed diffusion scaling has not yet been addressed.

TVAP adds a $\delta p$ diffusion term only to the momentum equations, so in effect uses the Liou-Steffen form for both the velocity and pressure perturbation diffusion terms, albeit with no $\delta U$ diffusion on the momentum equations.
TV-MAS keeps the Toro-Vasquez form of the $\delta p$ terms in the momentum and energy equations, but uses the mixed scaling for the $\mu_{22}$ coefficient in the momentum equations and the acoustic scaling in the energy equations.
Lin et al. \cite{lin_density_2018} show that having different $\mu_{22}$ scalings in each equation results in the density variations scaling as $\nabla\rho\sim\mathcalOM{}$, instead of the expected $\nabla\rho\sim\mathcalOM{2}$ for isentropic flow, even when the pressure variations are correct at $\nabla p\sim\mathcalOM{2}$.
They rectify this in TV-MAS2, and show that two low Mach number Roe-type schemes with the mixed diffusion scaling also require this correction \cite{thornber_numerical_2008,fillion_flica-ovap_2011}.\\

In this section we have discussed the diffusion scaling and form of a number of low Mach number convection-pressure flux-vector splitting schemes from the literature.
The Liou-Steffen splitting is the most popular, although schemes using the Zha-Bilgen and Toro-Vasquez splittings have also appeared recently.
All of the Liou-Steffen and Zha-Bilgen schemes surveyed match one of the general forms used here.
On the other hand, the Toro-Vasquez schemes do not match exactly, applying the velocity perturbation only to the density and energy equations, although the current analysis is presumed to still give a reasonable indication of their low Mach number behaviour.
Of the Zha-Bilgen and Toro-Vasquez schemes surveyed, every one has $\delta U$ diffusion equivalent to the Liou-Steffen and Roe forms on the energy equation.
The majority of schemes use the mixed diffusion scaling, especially in the more recent literature.
One notable exception to this is the AUSM$^+$-up scheme which uses the convective diffusion scaling, so should be treated as asymptotically equivalent to the preconditioned Roe scheme as $M\to0$.

\section{Numerical examples}\label{sec:numerical-examples}

Three numerical examples are presented to demonstrate the behaviour of the different splittings with both the convective and mixed diffusion scalings for convective and acoustic flows.
Results are shown for both the first and second forms of the Zha-Bilgen and Toro-Vasquez forms, with particular attention paid to the predicted entropy field for each example to show the effect of the erroneous diffusion terms found in section \ref{sec:cp-splittings}.

The same finite volume code used in \cite{hope-collins_artificial_2022} is used here, with the only difference being the interface flux functions.
These are constructed using (\ref{eq:general-cp-interface-flux}), (\ref{eq:general-interface-up}) and (\ref{eq:fds-interface-delta-up}).
$\underline{\varphi}$ is taken as the upwind value based on the sign of $U_{1/2}$, and the form of $\delta\underline{P}$ is as described for each splitting in section \ref{sec:cp-splittings-conserved}.
The scheme is first order in space, and uses first order Euler forward integration in time.
Note that this means that $1/M$ times as many timesteps are required for the convective scheme due to the increased spectral radius of the diffusion.
An ideal gas law is used for all cases, with $\gamma=1.4$ and $R=287.058$.

\subsection{One dimensional examples}

\subsubsection{Isolated soundwave}

\begin{figure}
    \centering
\begin{subfigure}[t]{0.49\textwidth}
    \centering
    \includegraphics[width=0.99\linewidth]{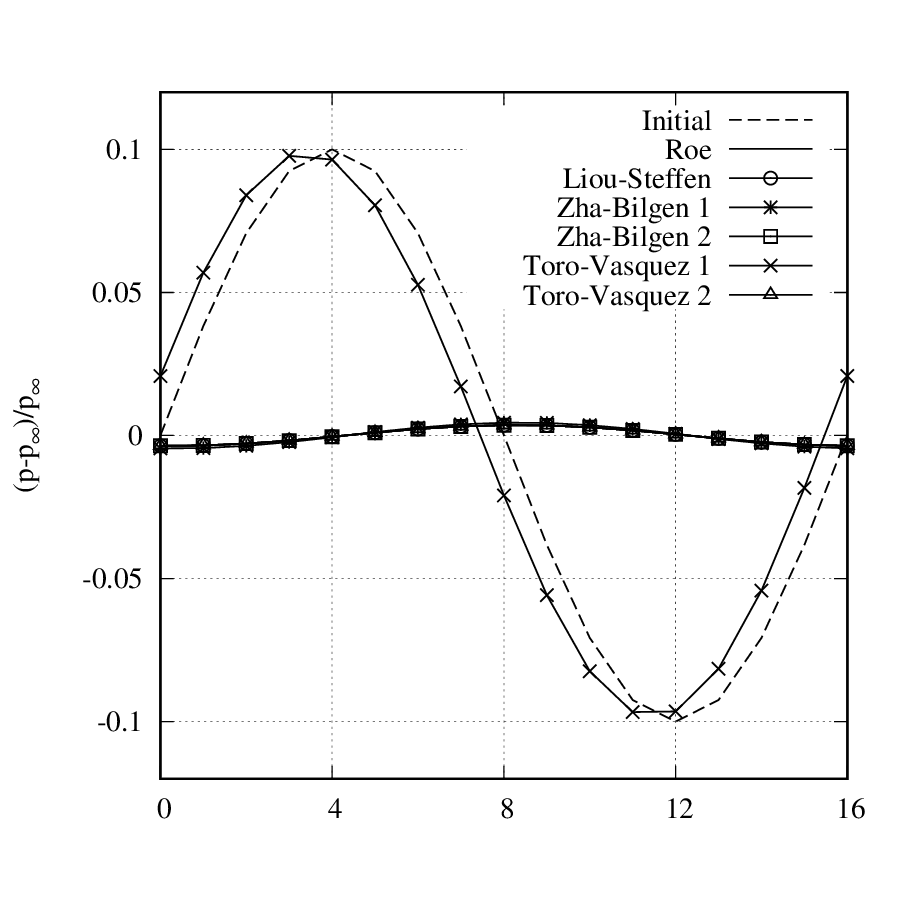}
    \caption{}
    \label{fig:soundwave-convective}
\end{subfigure}
\begin{subfigure}[t]{0.49\textwidth}
    \centering
    \includegraphics[width=0.99\linewidth]{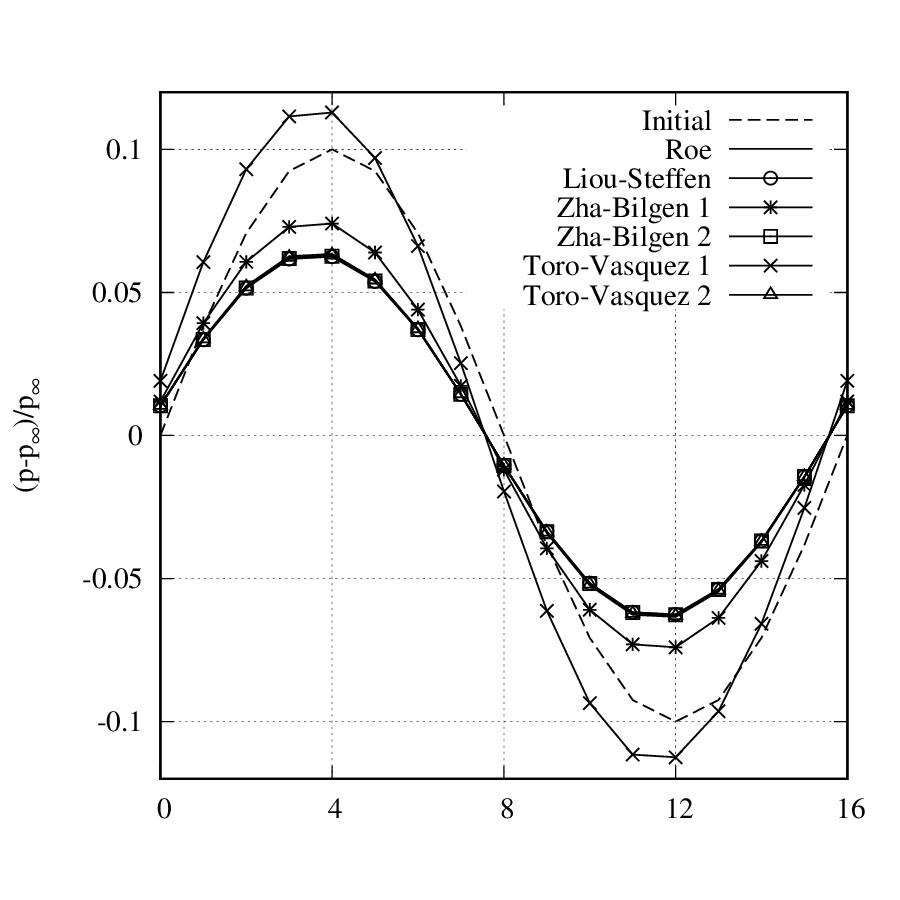}
    \caption{}
    \label{fig:soundwave-mixed}
\end{subfigure}
    \caption{Non-dimensional gauge pressure profiles for an isolated soundwave at $M=0.01$ after a single period using (a) convective diffusion scaling $\uuline{A}^c$ (b) mixed diffusion scaling $\uuline{A}^m$.}
    \label{fig:soundwave}
\end{figure}

The performance of the discrete schemes for purely acoustic flow is demonstrated using the isolated soundwave test case from \cite{hope-collins_artificial_2022}.
The initial profile, shown in figure \ref{fig:soundwave}, is a sinusoidal right travelling isentropic soundwave with a background flow at $M=0.01$ and 16 points per wavelength.
Pressure profiles after a single period are shown for each of the flux-splittings with the convective and mixed diffusion scaling in figures \ref{fig:soundwave-convective} and \ref{fig:soundwave-mixed} respectively, with the Roe scheme results shown for comparison.

With the convective diffusion scaling, all schemes except the first Toro-Vasquez form quickly damp the acoustic variations and closely match the results with the Roe scheme, as predicted by the limit equations (\ref{eq:limit-ls-Ac-acoustic}) and (\ref{eq:limit-zb-Ac-acoustic}).
The first form of the Toro-Vasquez scheme does not damp the acoustic wave at all due to the lack of pressure diffusion on the pressure equation.
The results with the mixed diffusion scaling, shown in figure \ref{fig:soundwave-mixed}, also agree with the findings of section \ref{sec:cp-splittings}.
The Liou-Steffen scheme and the second Zha-Bilgen and Toro-Vasquez forms match the Roe scheme almost exactly, and the first Zha-Bilgen form has similar resolution but slightly reduced diffusion.
The first Toro-Vasquez form predicts the soundwave amplitude growing, which is unsurprising given that the scheme asymptotically approaches a central scheme for acoustic flow, and a first order Euler forward discretisation is used for the time derivative.

\begin{figure}
    \centering
    \begin{subfigure}[t]{0.49\textwidth}
        \centering
        \includegraphics[width=0.99\linewidth]{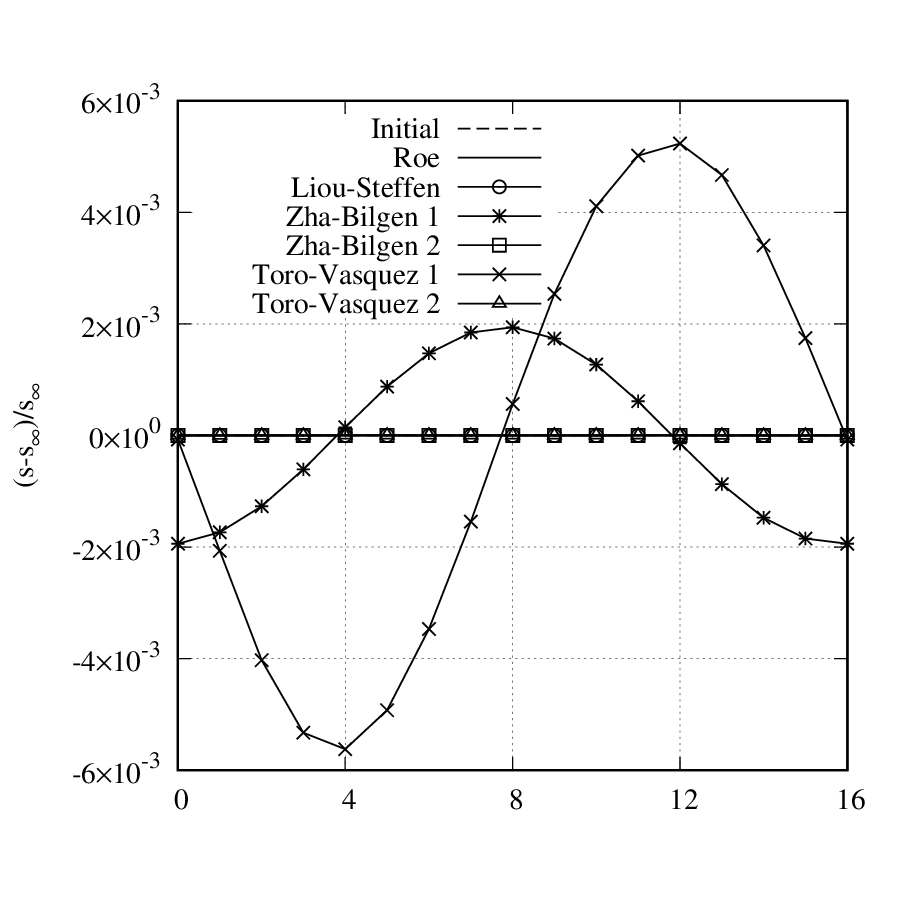}
        \caption{}
        \label{fig:soundwave-convective-entropy}
    \end{subfigure}
    \begin{subfigure}[t]{0.49\textwidth}
        \centering
        \includegraphics[width=0.99\linewidth]{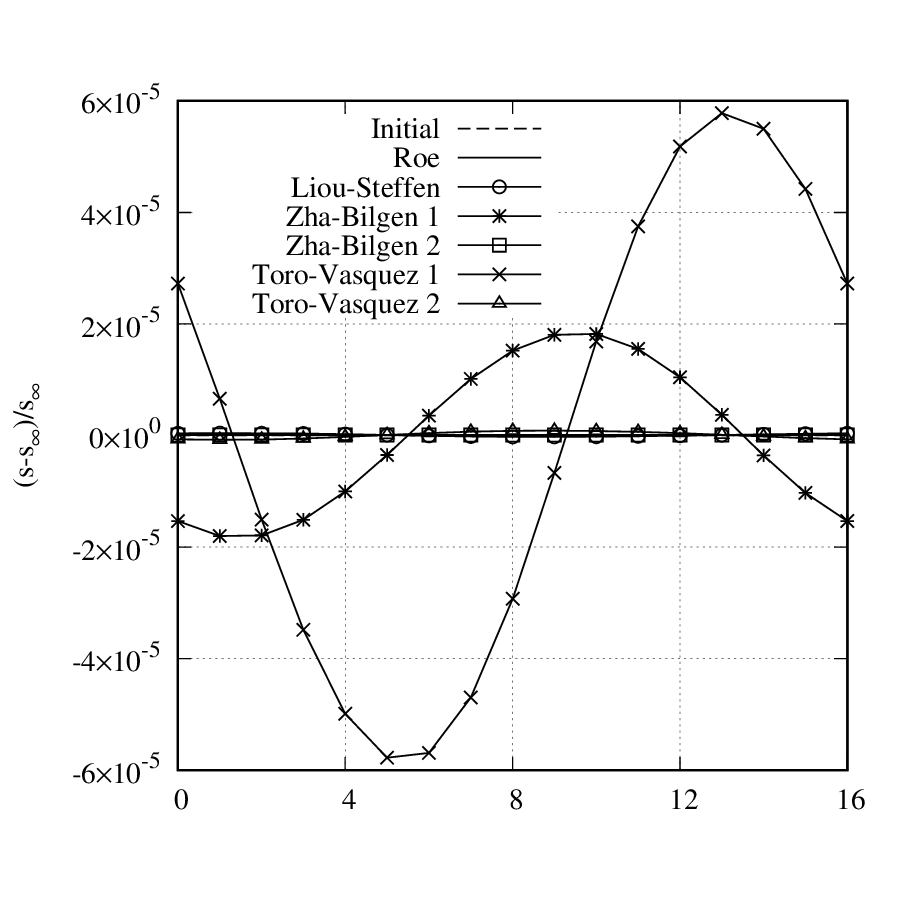}
        \caption{}
        \label{fig:soundwave-mixed-entropy}
    \end{subfigure}
    \caption{Non-dimensional entropy profiles for an isolated soundwave at $M=0.01$ after a single period using (a) convective diffusion scaling $\uuline{A}^c$ (b) mixed diffusion scaling $\uuline{A}^m$.}
    \label{fig:soundwave-entropy}
\end{figure}

Entropic profiles of $s=p/\rho^{\gamma}$ predicted by the different schemes are shown in figure \ref{fig:soundwave-entropy}.
The Roe and Liou-Steffen schemes preserve the isentropic nature of the soundwave for both convective and mixed diffusion scalings, as predicted by the limit equation analysis.
The second Zha-Bilgen and Toro-Vasquez forms preserve the isentropic profile to a similar degree, but have fractionally more entropy generation just visible in figure \ref{fig:soundwave-mixed-entropy}.
However, the first Zha-Bilgen and Toro-Vasquez forms both generate spurious entropy variations.
For the convective diffusion scaling, this can be seen from the limit equations (\ref{eq:limit-zb-Ac-acoustic}) and (\ref{eq:limit-tv-Ac-acoustic}) to be due to the erroneous pressure diffusion term in the acoustic entropy.
This term is a factor of $\gamma/(\gamma-1)$ larger in the Toro-Vasquez splitting and $\nabla^2p$ is larger because the soundwave is undamped, which results in a higher entropy generation for this scheme.
The variations are less than $1\%$ of the background value, but this is after the soundwave has travelled just a single wavelength.
With the mixed diffusion scaling, these erroneous pressure diffusion terms are one order of $M$ smaller than the leading order terms in the entropy equations of these two schemes, so disappear from the limit equations (\ref{eq:limit-zb-Am-acoustic}) and (\ref{eq:limit-tv-Am-acoustic}).
Figure \ref{fig:soundwave-mixed-entropy} shows that the entropy variations predicted with the mixed diffusion scaling are a factor of $M$ smaller than those predicted by the convective diffusion scaling, consistent with this scaling of the pressure diffusion terms.
The entropy profiles are essentially unchanged if the test case is rerun with the $\delta p$ terms in the interface pressure perturbation $\delta\underline{P}$ switched off, which further reinforces that the entropy generation of the first Zha-Bilgen and Toro-Vasquez forms is due to the form of the $\delta U$ diffusion terms in the energy equation.

\subsection{Two dimensional examples}

\subsubsection{Circular cylinder}
The next test case is inviscid flow at $M=0.01$ around a circular cylinder, which will demonstrate the performance of the different schemes for purely convective stationary flow.
The setup is identical to that in \cite{hope-collins_artificial_2022}.
A stretched O-grid is used with the first cell height of $0.036r$, where $r$ is the cylinder radius, and the farfield boundary is $50r$ from the centre of the cylinder.
The farfield boundary conditions are enforced by setting ghost cell values using upstream velocity and entropy and downstream static pressure.
At the cylinder wall, curvature corrected boundary conditions are used which reduce spurious entropy production at the wall \cite{dadone_surface_1994}.
Convergence to steady state is accelerated using local timestepping and Weiss \& Smith's low Mach number preconditioner \cite{weiss_preconditioning_1995}.

\begin{figure}
    \centering
    \begin{subfigure}[b]{0.475\textwidth}
        \centering
        \includegraphics[width=0.99\textwidth]{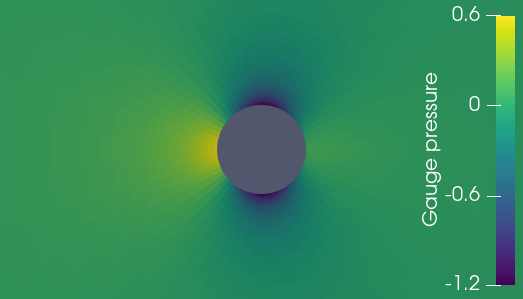}
        \caption{Gauge pressure}
        \label{fig:cylinder-roe-convective-pressure}
    \end{subfigure}
    \hfill
    \begin{subfigure}[b]{0.475\textwidth}
        \centering
        \includegraphics[width=0.99\textwidth]{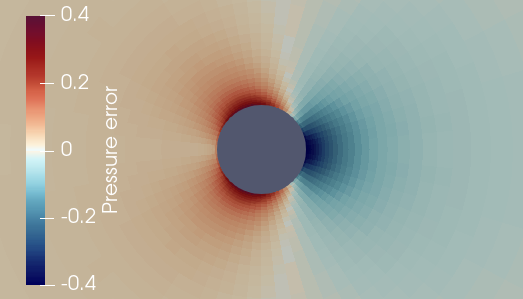}
        \caption{Pressure error}
        \label{fig:cylinder-roe-convective-pressure-error}
    \end{subfigure}
    \vskip\baselineskip
    \begin{subfigure}[b]{0.475\textwidth}
        \centering
        \includegraphics[width=0.99\textwidth]{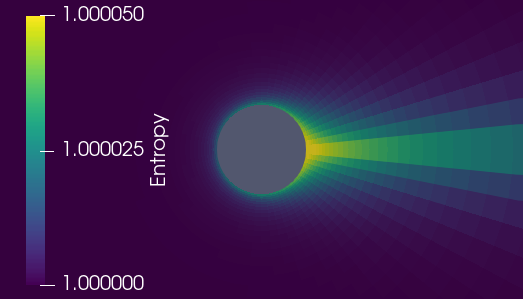}
        \caption{Entropy}
        \label{fig:cylinder-roe-convective-entropy}
    \end{subfigure}
    \hfill
    \begin{subfigure}[b]{0.475\textwidth}
        \centering
        \includegraphics[width=0.99\textwidth]{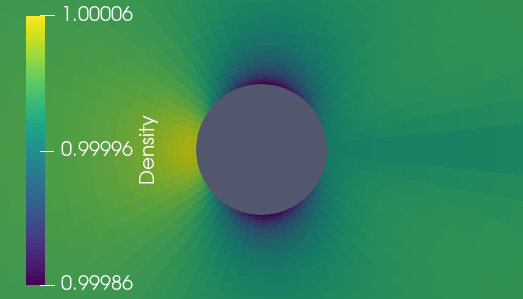}
        \caption{Density}
        \label{fig:cylinder-roe-convective-density}
    \end{subfigure}
    \caption{Flow around a circular cylinder at $M=0.01$ using the Roe scheme with convective diffusion scaling.}
    \label{fig:cylinder-roe-convective}
\end{figure}

\begin{figure}
    \centering
    \begin{subfigure}[b]{0.475\textwidth}
        \centering
        \includegraphics[width=0.99\textwidth]{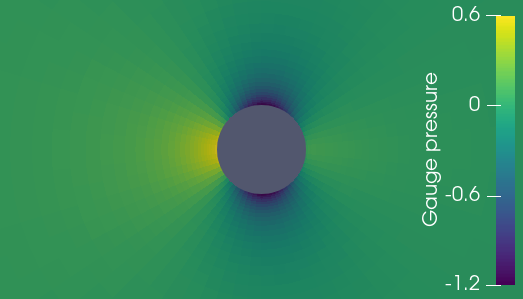}
        \caption{Gauge pressure}
        \label{fig:cylinder-ls-convective-pressure}
    \end{subfigure}
    \hfill
    \begin{subfigure}[b]{0.475\textwidth}
        \centering
        \includegraphics[width=0.99\textwidth]{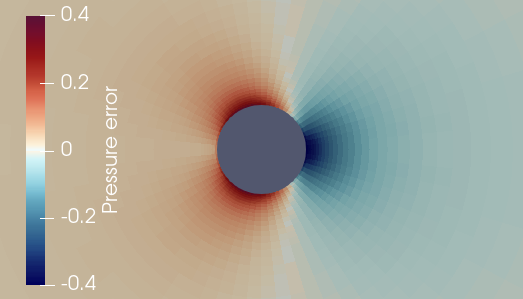}
        \caption{Pressure error}
        \label{fig:cylinder-ls-convective-pressure-error}
    \end{subfigure}
    \vskip\baselineskip
    \begin{subfigure}[b]{0.475\textwidth}
        \centering
        \includegraphics[width=0.99\textwidth]{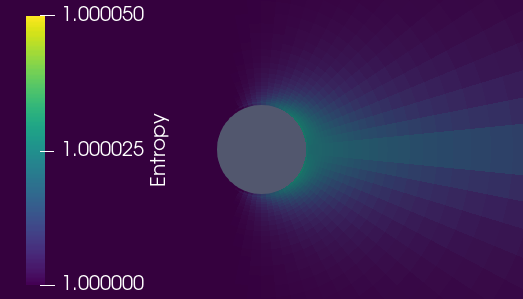}
        \caption{Entropy}
        \label{fig:cylinder-ls-convective-entropy}
    \end{subfigure}
    \hfill
    \begin{subfigure}[b]{0.475\textwidth}
        \centering
        \includegraphics[width=0.99\textwidth]{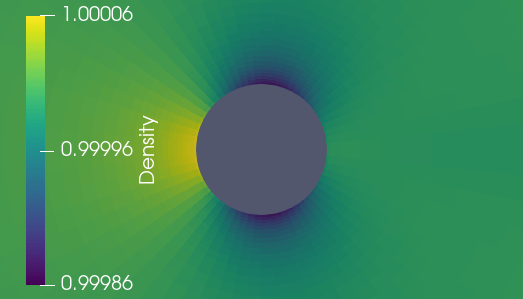}
        \caption{Density}
        \label{fig:cylinder-ls-convective-density}
    \end{subfigure}
    \caption{Flow around a circular cylinder at $M=0.01$ using the Liou-Steffen scheme with convective diffusion scaling.}
    \label{fig:cylinder-ls-convective}
\end{figure}

The profiles of the gauge pressure, the pressure error relative to the analytical potential flow solution, the entropy, and the density using the Roe and Liou-Steffen schemes with the convective artificial diffusion scaling are shown in figures \ref{fig:cylinder-roe-convective} and \ref{fig:cylinder-ls-convective}.
As for the isolated soundwave, the Liou-Steffen results are almost indistinguishable from the Roe results.
The numerical results are close to the potential solution, with the pressure variations \ref{fig:cylinder-roe-convective-pressure} and \ref{fig:cylinder-ls-convective-pressure} on the correct order of magnitude.
The pressure loss downstream of the cylinder is due to the fact a first order scheme has been used, but the smooth error profiles \ref{fig:cylinder-roe-convective-pressure-error} and \ref{fig:cylinder-ls-convective-pressure-error} indicate that the pressure is free of chequerboard modes.
There is a very small amount of entropy generated at the wall visible in \ref{fig:cylinder-roe-convective-entropy} and \ref{fig:cylinder-ls-convective-entropy} - which is unavoidable with standard upwind schemes, although is reduced by the curvature corrected boundary conditions (see \cite{gouasmi_entropy-stable_2022} for a recent example of low Mach number entropy conserving/stable schemes) - and the density profiles \ref{fig:cylinder-roe-convective-density} and \ref{fig:cylinder-ls-convective-density} resemble the pressure profiles as expected.

\begin{figure}
    \centering
    \begin{subfigure}[b]{0.475\textwidth}
        \centering
        \includegraphics[width=0.99\textwidth]{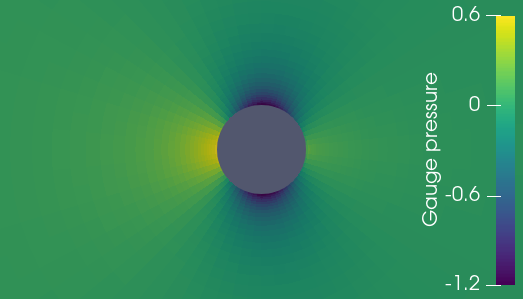}
        \caption{Gauge pressure}
        \label{fig:cylinder-zb-convective-pressure1}
    \end{subfigure}
    \hfill
    \begin{subfigure}[b]{0.475\textwidth}
        \centering
        \includegraphics[width=0.99\textwidth]{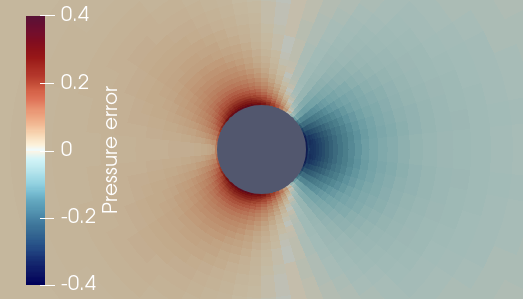}
        \caption{Pressure error}
        \label{fig:cylinder-zb-convective-pressure-error1}
    \end{subfigure}
    \vskip\baselineskip
    \begin{subfigure}[b]{0.475\textwidth}
        \centering
        \includegraphics[width=0.99\textwidth]{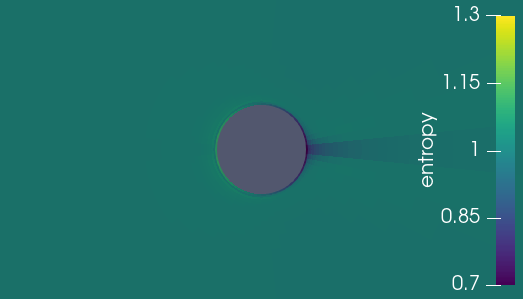}
        \caption{Entropy}
        \label{fig:cylinder-zb-convective-entropy1}
    \end{subfigure}
    \hfill
    \begin{subfigure}[b]{0.475\textwidth}
        \centering
        \includegraphics[width=0.99\textwidth]{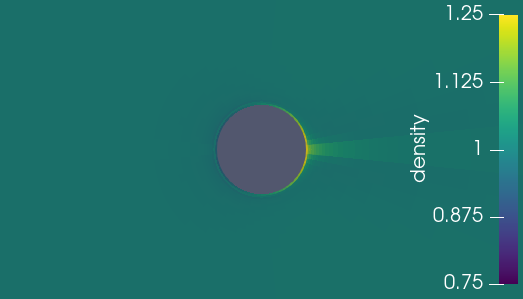}
        \caption{Density}
        \label{fig:cylinder-zb-convective-density1}
    \end{subfigure}
    \caption{Flow around a circular cylinder at $M=0.01$ using the first Zha-Bilgen scheme with convective diffusion scaling.}
    \label{fig:cylinder-zb-convective1}
\end{figure}

\begin{figure}
    \centering
    \begin{subfigure}[b]{0.475\textwidth}
        \centering
        \includegraphics[width=0.99\textwidth]{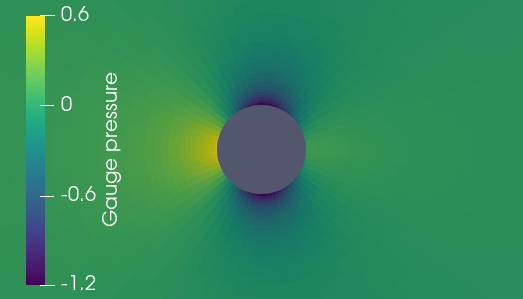}
        \caption{Gauge pressure}
        \label{fig:cylinder-zb-convective-pressure2}
    \end{subfigure}
    \hfill
    \begin{subfigure}[b]{0.475\textwidth}
        \centering
        \includegraphics[width=0.99\textwidth]{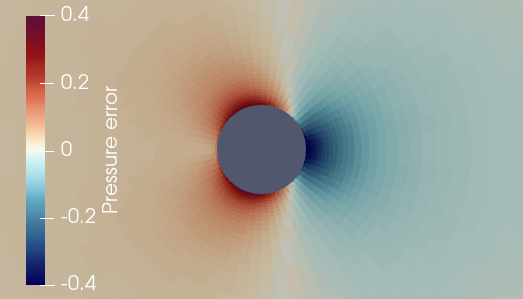}
        \caption{Pressure error}
        \label{fig:cylinder-zb-convective-pressure-error2}
    \end{subfigure}
    \vskip\baselineskip
    \begin{subfigure}[b]{0.475\textwidth}
        \centering
        \includegraphics[width=0.99\textwidth]{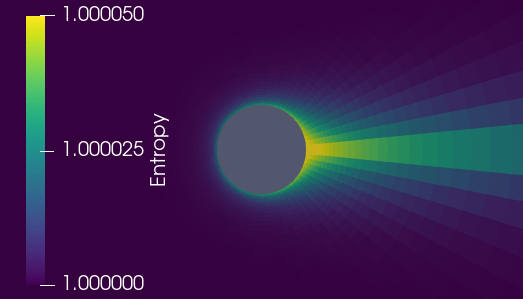}
        \caption{Entropy}
        \label{fig:cylinder-zb-convective-entropy2}
    \end{subfigure}
    \hfill
    \begin{subfigure}[b]{0.475\textwidth}
        \centering
        \includegraphics[width=0.99\textwidth]{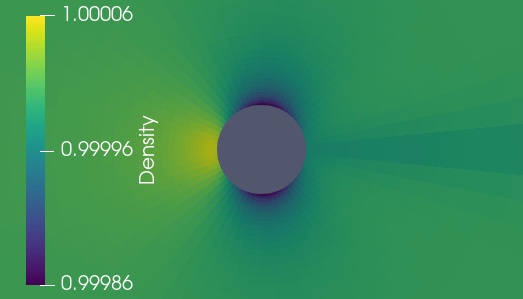}
        \caption{Density}
        \label{fig:cylinder-zb-convective-density2}
    \end{subfigure}
    \caption{Flow around a circular cylinder at $M=0.01$ using the second Zha-Bilgen scheme with convective diffusion scaling.}
    \label{fig:cylinder-zb-convective2}
\end{figure}

The results for the first Zha-Bilgen splitting with the convective diffusion scaling are shown in figure \ref{fig:cylinder-zb-convective1}.
The pressure field \ref{fig:cylinder-zb-convective-pressure1} is essentially the same as that of the Roe and Liou-Steffen schemes, although the error plot \ref{fig:cylinder-zb-convective-pressure-error1} shows that the pressure loss downstream of the cylinder is marginally less due to the reduced diffusion in the  pressure equation.
However, the entropy profile \ref{fig:cylinder-zb-convective-entropy1} shows large unphysical variations due to the pressure anti-diffusion term in the entropy equation, which peak at around $30\%$ of the freestream values close to the wall where the pressure Poisson $\nabla^2p$ is largest.
The effect of this entropy generation can be seen in the density profile \ref{fig:cylinder-zb-convective-density1}, where large variations can also be seen close to the wall.
The results for the second Zha-Bilgen splitting with the convective diffusion scaling are shown in figure \ref{fig:cylinder-zb-convective2}.
The pressure and pressure error fields are essentially indistinguishable from the first form, but the entropy and density fields show clear improvement.
The entropy variations have been reduced to approximately the same level as the Roe and Liou-Steffen schemes, and the density profile is vastly improved.

\begin{figure}
    \centering
    \begin{subfigure}[b]{0.475\textwidth}
        \centering
        \includegraphics[width=0.99\textwidth]{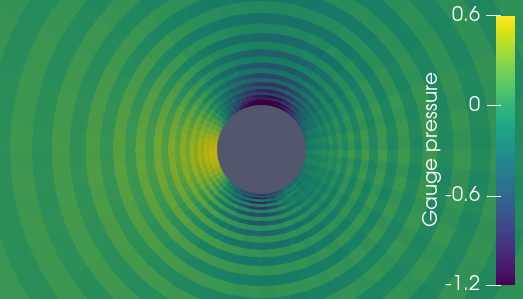}
        \caption{Gauge pressure}
        \label{fig:cylinder-tv-convective-pressure1}
    \end{subfigure}
    \hfill
    \begin{subfigure}[b]{0.475\textwidth}
        \centering
        \includegraphics[width=0.99\textwidth]{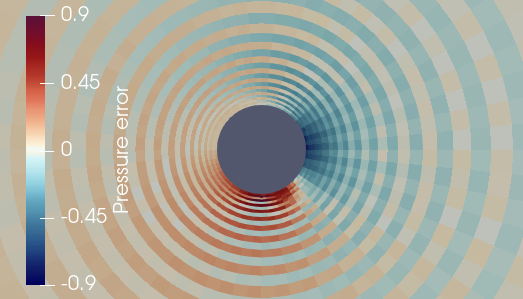}
        \caption{Pressure error}
        \label{fig:cylinder-tv-convective-pressure-error1}
    \end{subfigure}
    \vskip\baselineskip
    \begin{subfigure}[b]{0.475\textwidth}
        \centering
        \includegraphics[width=0.99\textwidth]{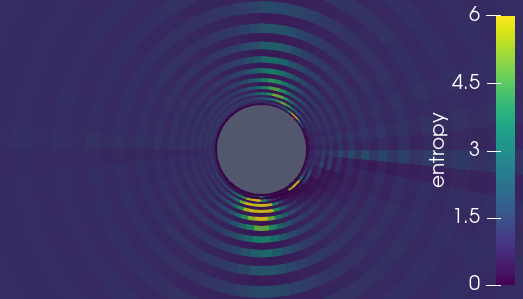}
        \caption{Entropy}
        \label{fig:cylinder-tv-convective-entropy1}
    \end{subfigure}
    \hfill
    \begin{subfigure}[b]{0.475\textwidth}
        \centering
        \includegraphics[width=0.99\textwidth]{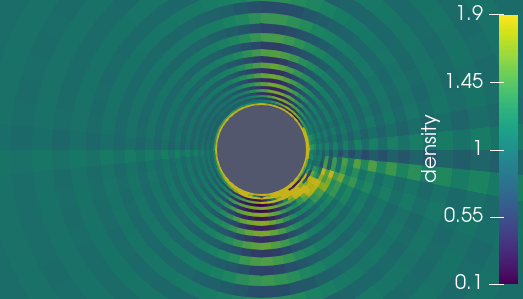}
        \caption{Density}
        \label{fig:cylinder-tv-convective-density1}
    \end{subfigure}
    \caption{Flow around a circular cylinder at $M=0.01$ using the first Toro-Vasquez scheme with convective diffusion scaling.}
    \label{fig:cylinder-tv-convective1}
\end{figure}

\begin{figure}
    \centering
    \begin{subfigure}[b]{0.475\textwidth}
        \centering
        \includegraphics[width=0.99\textwidth]{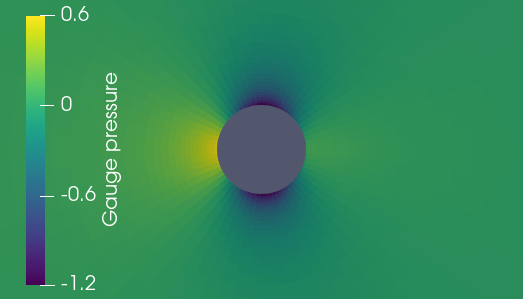}
        \caption{Gauge pressure}
        \label{fig:cylinder-tv-convective-pressure2}
    \end{subfigure}
    \hfill
    \begin{subfigure}[b]{0.475\textwidth}
        \centering
        \includegraphics[width=0.99\textwidth]{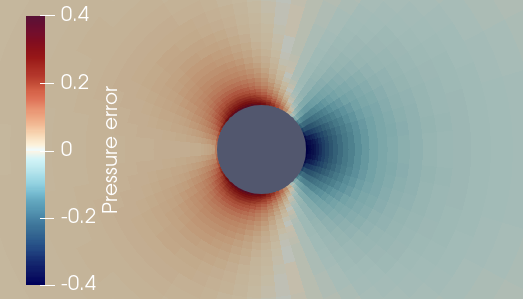}
        \caption{Pressure error}
        \label{fig:cylinder-tv-convective-pressure-error2}
    \end{subfigure}
    \vskip\baselineskip
    \begin{subfigure}[b]{0.475\textwidth}
        \centering
        \includegraphics[width=0.99\textwidth]{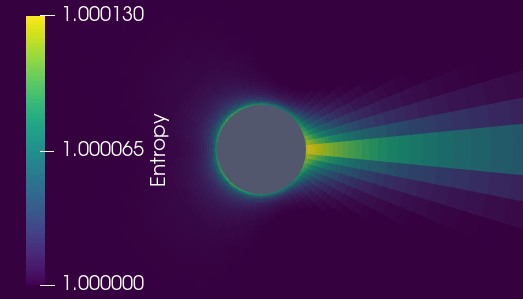}
        \caption{Entropy}
        \label{fig:cylinder-tv-convective-entropy2}
    \end{subfigure}
    \hfill
    \begin{subfigure}[b]{0.475\textwidth}
        \centering
        \includegraphics[width=0.99\textwidth]{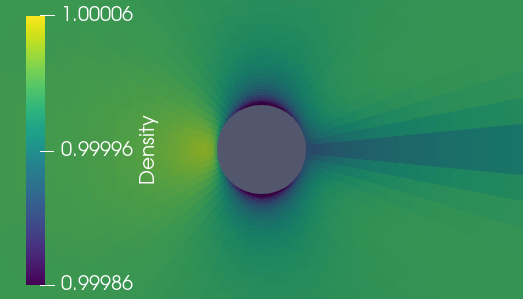}
        \caption{Density}
        \label{fig:cylinder-tv-convective-density2}
    \end{subfigure}
    \caption{Flow around a circular cylinder at $M=0.01$ using the second Toro-Vasquez scheme with convective diffusion scaling.}
    \label{fig:cylinder-tv-convective2}
\end{figure}

The results for the first Toro-Vasquez form with the convective diffusion scaling are shown in figure \ref{fig:cylinder-tv-convective1}.
For this case, the previous schemes converged in around $25,000-30,000$ iterations, however the first Toro-Vasquez form did not converge, with NaNs occuring in the solution after around $42,000$ iterations.
Results are shown here after $30,000$ iterations.
Significant radial chequerboard modes can be seen in the pressure and pressure error profiles \ref{fig:cylinder-tv-convective-pressure1} and \ref{fig:cylinder-tv-convective-pressure-error1}.
The density and entropy profiles \ref{fig:cylinder-tv-convective-entropy1} and \ref{fig:cylinder-tv-convective-density1} also have radial chequerboard modes and large errors close to the wall due to the pressure anti-diffusion in the entropy equation.
These errors grow until the solution eventually diverges and breaks down.
The major difference between the limit equations of the first Zha-Bilgen and Toro-Vasquez forms with the convective diffusion scaling is the diffusion term in the pressure equation which, by comparing figures \ref{fig:cylinder-zb-convective1} and \ref{fig:cylinder-tv-convective1}, implies that this term somewhat damps the errors arising from the spurious entropy generation.
The results for the second Toro-Vasquez form with the convective diffusion scaling are shown in figure \ref{fig:cylinder-tv-convective2}.
As for the Zha-Bilgen scheme, the entropy and density fields show a marked improvement on the first form, although the entropy variations are about twice that seen for the Roe and Liou-Steffen schemes.\\

\begin{figure}
    \centering
    \begin{subfigure}[b]{0.475\textwidth}
        \centering
        \includegraphics[width=0.99\textwidth]{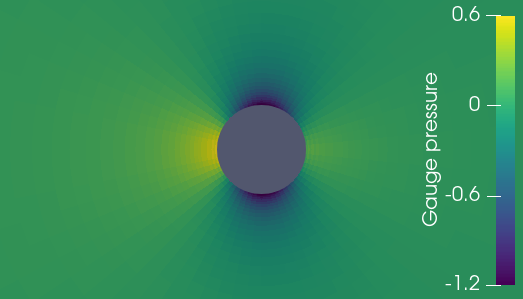}
        \caption{Gauge pressure}
        \label{fig:cylinder-roe-mixed-pressure}
    \end{subfigure}
    \hfill
    \begin{subfigure}[b]{0.475\textwidth}
        \centering
        \includegraphics[width=0.99\textwidth]{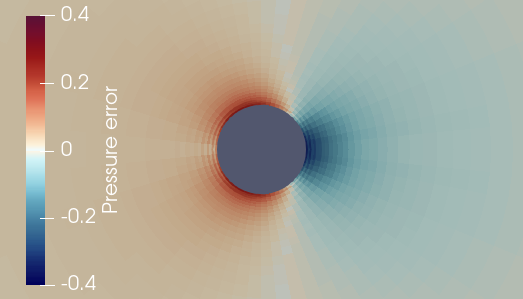}
        \caption{Pressure error}
        \label{fig:cylinder-roe-mixed-pressure-error}
    \end{subfigure}
    \vskip\baselineskip
    \begin{subfigure}[b]{0.475\textwidth}
        \centering
        \includegraphics[width=0.99\textwidth]{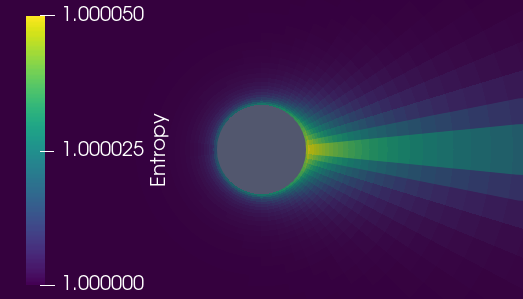}
        \caption{Entropy}
        \label{fig:cylinder-roe-mixed-entropy}
    \end{subfigure}
    \hfill
    \begin{subfigure}[b]{0.475\textwidth}
        \centering
        \includegraphics[width=0.99\textwidth]{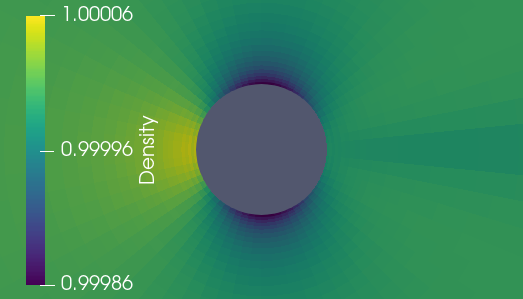}
        \caption{Density}
        \label{fig:cylinder-roe-mixed-density}
    \end{subfigure}
    \caption{Flow around a circular cylinder at $M=0.01$ using the Roe scheme with mixed diffusion scaling.}
    \label{fig:cylinder-roe-mixed}
\end{figure}

\begin{figure}
    \centering
    \begin{subfigure}[b]{0.475\textwidth}
        \centering
        \includegraphics[width=0.99\textwidth]{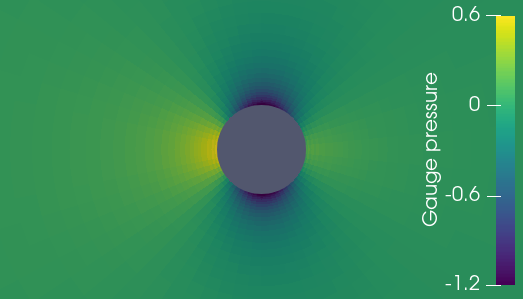}
        \caption{Gauge pressure}
        \label{fig:cylinder-ls-mixed-pressure}
    \end{subfigure}
    \hfill
    \begin{subfigure}[b]{0.475\textwidth}
        \centering
        \includegraphics[width=0.99\textwidth]{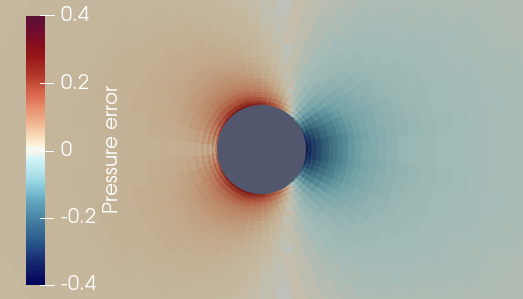}
        \caption{Pressure error}
        \label{fig:cylinder-ls-mixed-pressure-error}
    \end{subfigure}
    \vskip\baselineskip
    \begin{subfigure}[b]{0.475\textwidth}
        \centering
        \includegraphics[width=0.99\textwidth]{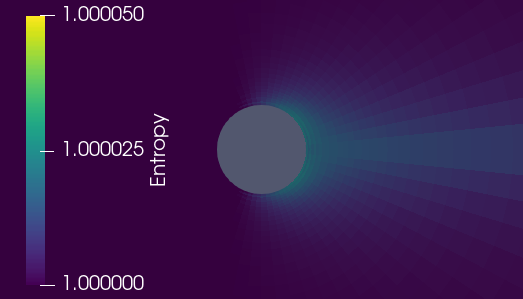}
        \caption{Entropy}
        \label{fig:cylinder-ls-mixed-entropy}
    \end{subfigure}
    \hfill
    \begin{subfigure}[b]{0.475\textwidth}
        \centering
        \includegraphics[width=0.99\textwidth]{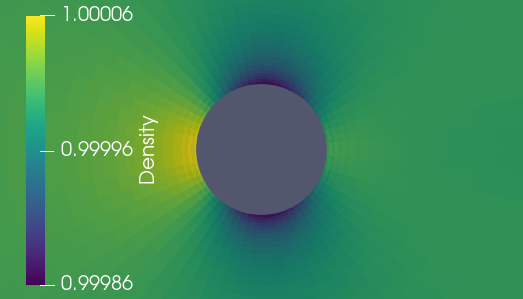}
        \caption{Density}
        \label{fig:cylinder-ls-mixed-density}
    \end{subfigure}
    \caption{Flow around a circular cylinder at $M=0.01$ using the Liou-Steffen scheme with mixed diffusion scaling.}
    \label{fig:cylinder-ls-mixed}
\end{figure}

The results for the Roe and Liou-Steffen schemes with the mixed diffusion scaling are shown in figures \ref{fig:cylinder-roe-mixed} and \ref{fig:cylinder-ls-mixed} respectively.
The pressure and density plots are very similar to the results found with the convective diffusion scaling, and the entropy generated at the walls is even lower.
Inspection of figures \ref{fig:cylinder-roe-mixed-pressure-error} and \ref{fig:cylinder-ls-mixed-pressure-error} reveals that the error profiles are no longer smooth, and a slight chequerboard exists in both solutions, as is expected for schemes with the mixed diffusion scaling.

\begin{figure}
    \centering
    \begin{subfigure}[b]{0.475\textwidth}
        \centering
        \includegraphics[width=0.99\textwidth]{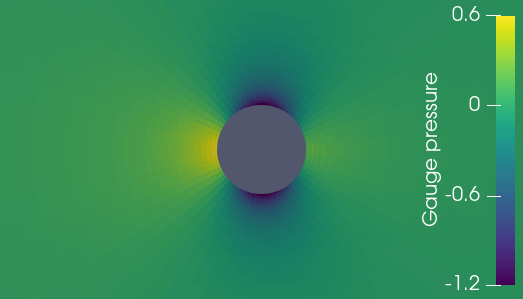}
        \caption{Gauge pressure}
        \label{fig:cylinder-zb-mixed-pressure1}
    \end{subfigure}
    \hfill
    \begin{subfigure}[b]{0.475\textwidth}
        \centering
        \includegraphics[width=0.99\textwidth]{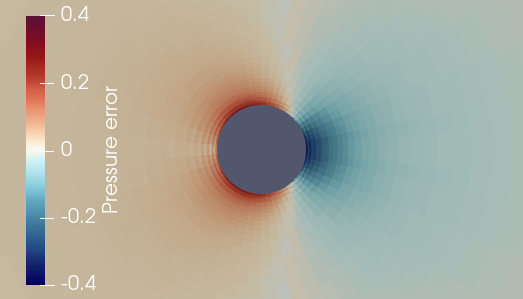}
        \caption{Pressure error}
        \label{fig:cylinder-zb-mixed-pressure-error1}
    \end{subfigure}
    \vskip\baselineskip
    \begin{subfigure}[b]{0.475\textwidth}
        \centering
        \includegraphics[width=0.99\textwidth]{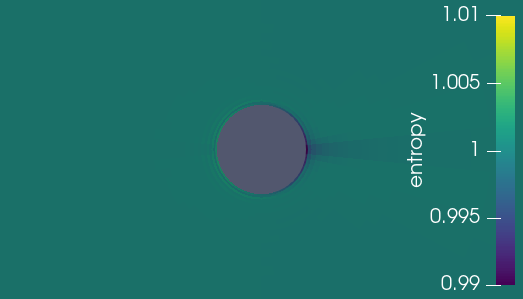}
        \caption{Entropy}
        \label{fig:cylinder-zb-mixed-entropy1}
    \end{subfigure}
    \hfill
    \begin{subfigure}[b]{0.475\textwidth}
        \centering
        \includegraphics[width=0.99\textwidth]{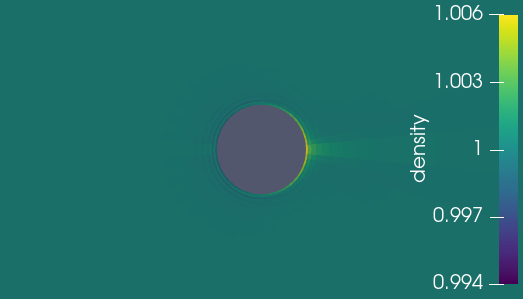}
        \caption{Density}
        \label{fig:cylinder-zb-mixed-density1}
    \end{subfigure}
    \caption{Flow around a circular cylinder at $M=0.01$ using the first Zha-Bilgen scheme with mixed diffusion scaling.}
    \label{fig:cylinder-zb-mixed1}
\end{figure}

\begin{figure}
    \centering
    \begin{subfigure}[b]{0.475\textwidth}
        \centering
        \includegraphics[width=0.99\textwidth]{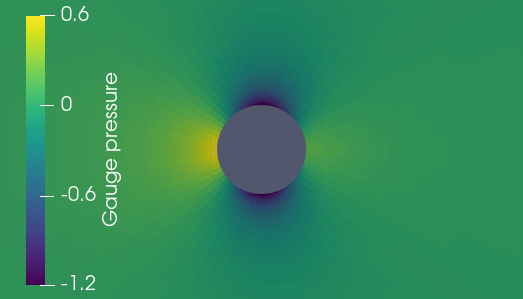}
        \caption{Gauge pressure}
        \label{fig:cylinder-zb-mixed-pressure2}
    \end{subfigure}
    \hfill
    \begin{subfigure}[b]{0.475\textwidth}
        \centering
        \includegraphics[width=0.99\textwidth]{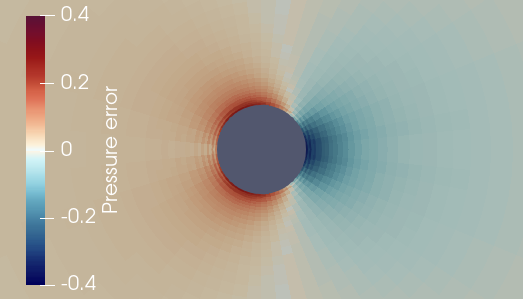}
        \caption{Pressure error}
        \label{fig:cylinder-zb-mixed-pressure-error2}
    \end{subfigure}
    \vskip\baselineskip
    \begin{subfigure}[b]{0.475\textwidth}
        \centering
        \includegraphics[width=0.99\textwidth]{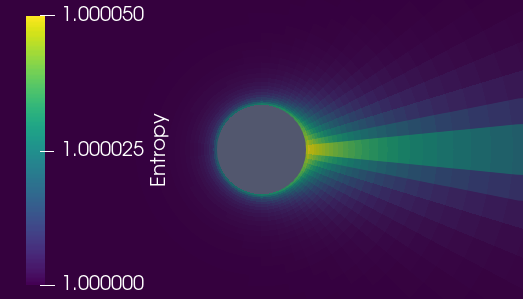}
        \caption{Entropy}
        \label{fig:cylinder-zb-mixed-entropy2}
    \end{subfigure}
    \hfill
    \begin{subfigure}[b]{0.475\textwidth}
        \centering
        \includegraphics[width=0.99\textwidth]{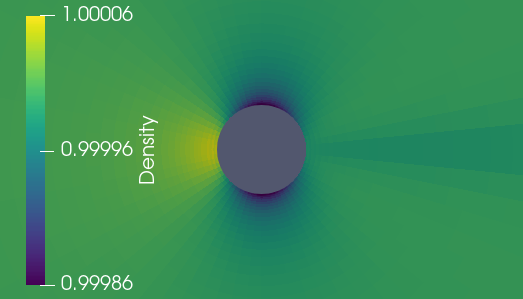}
        \caption{Density}
        \label{fig:cylinder-zb-mixed-density2}
    \end{subfigure}
    \caption{Flow around a circular cylinder at $M=0.01$ using the second Zha-Bilgen scheme with mixed diffusion scaling.}
    \label{fig:cylinder-zb-mixed2}
\end{figure}

The results for the first Zha-Bilgen form in figure \ref{fig:cylinder-zb-mixed1} with the mixed diffusion scaling also show that the pressure field is generally accurate, but a chequerboard mode also exists in the solution which is visible in \ref{fig:cylinder-zb-mixed-pressure-error1}.
The entropy and density variations \ref{fig:cylinder-zb-mixed-entropy1} and \ref{fig:cylinder-zb-mixed-density1} are approximately a factor of $M$ smaller than for the convective scaling \ref{fig:cylinder-zb-convective-entropy1} and \ref{fig:cylinder-zb-convective-density1},  although are still much larger than those found with the Roe and Liou-Steffen schemes.
The results for the second Zha-Bilgen form are shown in figure \ref{fig:cylinder-zb-mixed2} and closely resemble those of the Roe and Liou-Steffen schemes, including the pressure chequerboard and the level of the entropy variations.

\begin{figure}
    \centering
    \begin{subfigure}[b]{0.475\textwidth}
        \centering
        \includegraphics[width=0.99\textwidth]{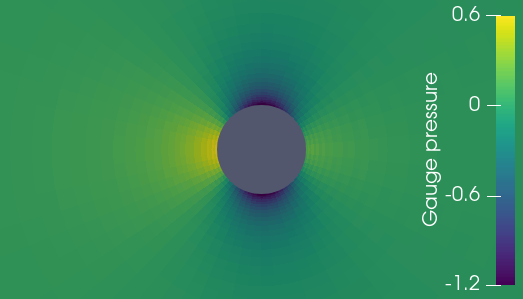}
        \caption{Gauge pressure}
        \label{fig:cylinder-tv-mixed-pressure1}
    \end{subfigure}
    \hfill
    \begin{subfigure}[b]{0.475\textwidth}
        \centering
        \includegraphics[width=0.99\textwidth]{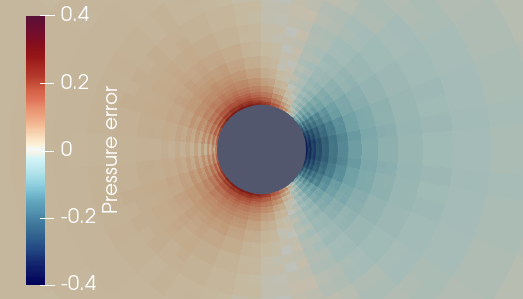}
        \caption{Pressure error}
        \label{fig:cylinder-tv-mixed-pressure-error1}
    \end{subfigure}
    \vskip\baselineskip
    \begin{subfigure}[b]{0.475\textwidth}
        \centering
        \includegraphics[width=0.99\textwidth]{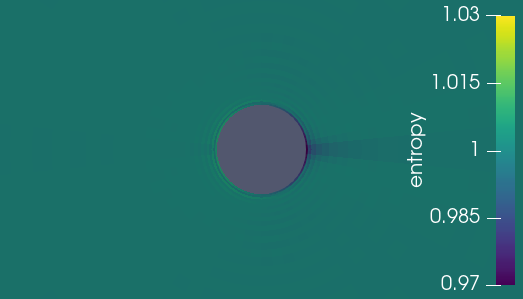}
        \caption{Entropy}
        \label{fig:cylinder-tv-mixed-entropy1}
    \end{subfigure}
    \hfill
    \begin{subfigure}[b]{0.475\textwidth}
        \centering
        \includegraphics[width=0.99\textwidth]{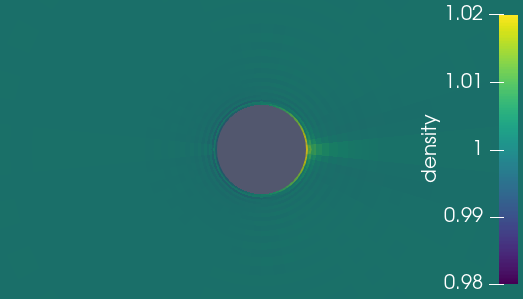}
        \caption{Density}
        \label{fig:cylinder-tv-mixed-density1}
    \end{subfigure}
    \caption{Flow around a circular cylinder at $M=0.01$ using the first Toro-Vasquez scheme with mixed diffusion scaling.}
    \label{fig:cylinder-tv-mixed1}
\end{figure}

\begin{figure}
    \centering
    \begin{subfigure}[b]{0.475\textwidth}
        \centering
        \includegraphics[width=0.99\textwidth]{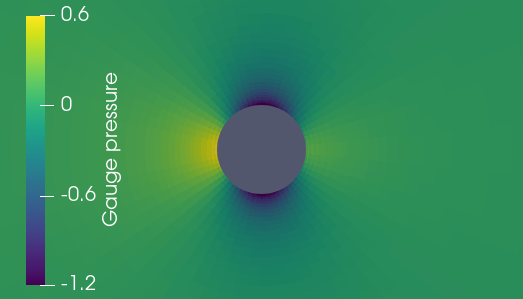}
        \caption{Gauge pressure}
        \label{fig:cylinder-tv-mixed-pressure2}
    \end{subfigure}
    \hfill
    \begin{subfigure}[b]{0.475\textwidth}
        \centering
        \includegraphics[width=0.99\textwidth]{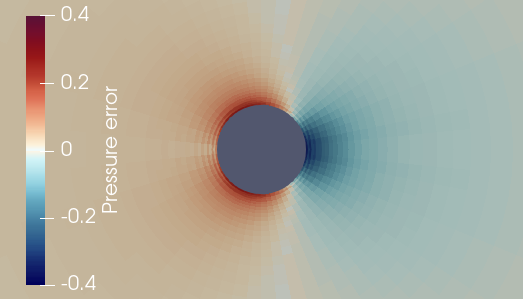}
        \caption{Pressure error}
        \label{fig:cylinder-tv-mixed-pressure-error2}
    \end{subfigure}
    \vskip\baselineskip
    \begin{subfigure}[b]{0.475\textwidth}
        \centering
        \includegraphics[width=0.99\textwidth]{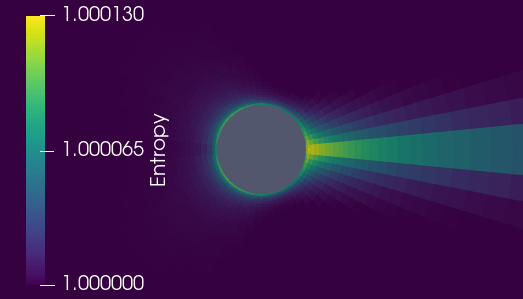}
        \caption{Entropy}
        \label{fig:cylinder-tv-mixed-entropy2}
    \end{subfigure}
    \hfill
    \begin{subfigure}[b]{0.475\textwidth}
        \centering
        \includegraphics[width=0.99\textwidth]{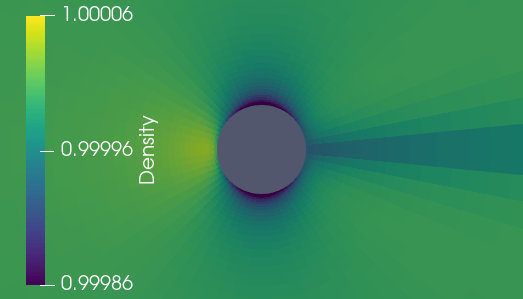}
        \caption{Density}
        \label{fig:cylinder-tv-mixed-density2}
    \end{subfigure}
    \caption{Flow around a circular cylinder at $M=0.01$ using the second Toro-Vasquez scheme with mixed diffusion scaling.}
    \label{fig:cylinder-tv-mixed2}
\end{figure}

The results for the first Toro-Vasquez form with mixed diffusion scaling in figure \ref{fig:cylinder-tv-mixed1} show a vast improvement on the results with the convective scaling in figure \ref{fig:cylinder-tv-convective1}.
The pressure field much more closely matches the potential solution, although still with a chequerboard mode visible in \ref{fig:cylinder-tv-mixed-pressure-error1}.
As for the Zha-Bilgen scheme, the entropy and density variations are much reduced compared to the convective scaling results - especially the entropy variations - while still being larger than the Roe and Liou-Steffen solutions.
The results for the second Toro-Vasquez form with mixed diffusion scaling are shown in figure \ref{fig:cylinder-tv-mixed2}.
As for with the convective diffusion scaling, the results are a significant improvement on the first form, although the entropy generation is still around twice that of the Roe and Liou-Steffen schemes.

\subsubsection{Isentropic vortex}

The final test case is an inviscid stationary Gresho vortex \cite{gresho_theory_1990}, which will be used to further demonstrate the impact of the form of the entropy limit equations.
The maximum velocity of the vortex $u_m$ is at a radius $r$ and Mach number $M=0.01$.
The initial conditions can be found in the soundwave-vortex interaction test case of \cite{hope-collins_artificial_2022}, with the exception that in the present study the density profile is calculated using $d\rho=dp/a^2$ to enforce isentropic conditions instead of using a constant value.
The domain is $10r\times10r$, and is periodic in both $x$ and $y$.
Results are shown after a quarter turn of the vortex, i.e. $t=\frac{\pi r}{2u_m}$.

\begin{figure}
    \centering
\begin{subfigure}[t]{0.49\textwidth}
    \centering
    \includegraphics[width=0.99\linewidth]{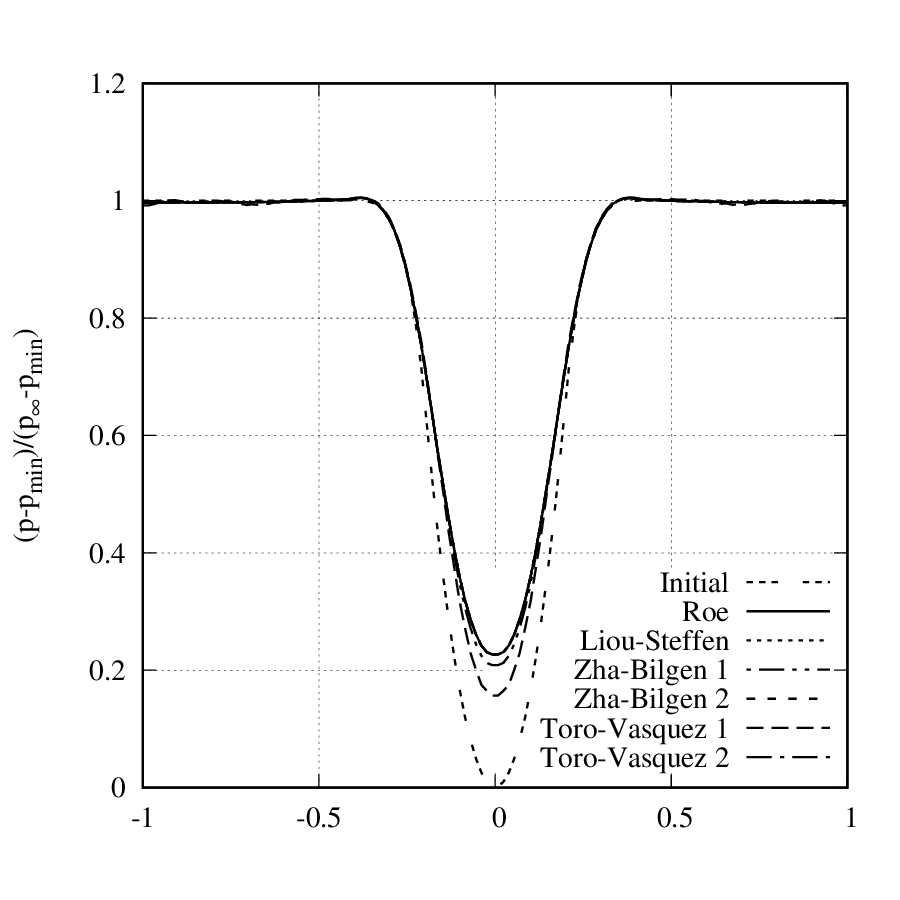}
    \caption{}
    \label{fig:vortex-convective-pressure}
\end{subfigure}
\begin{subfigure}[t]{0.49\textwidth}
    \centering
    \includegraphics[width=0.99\linewidth]{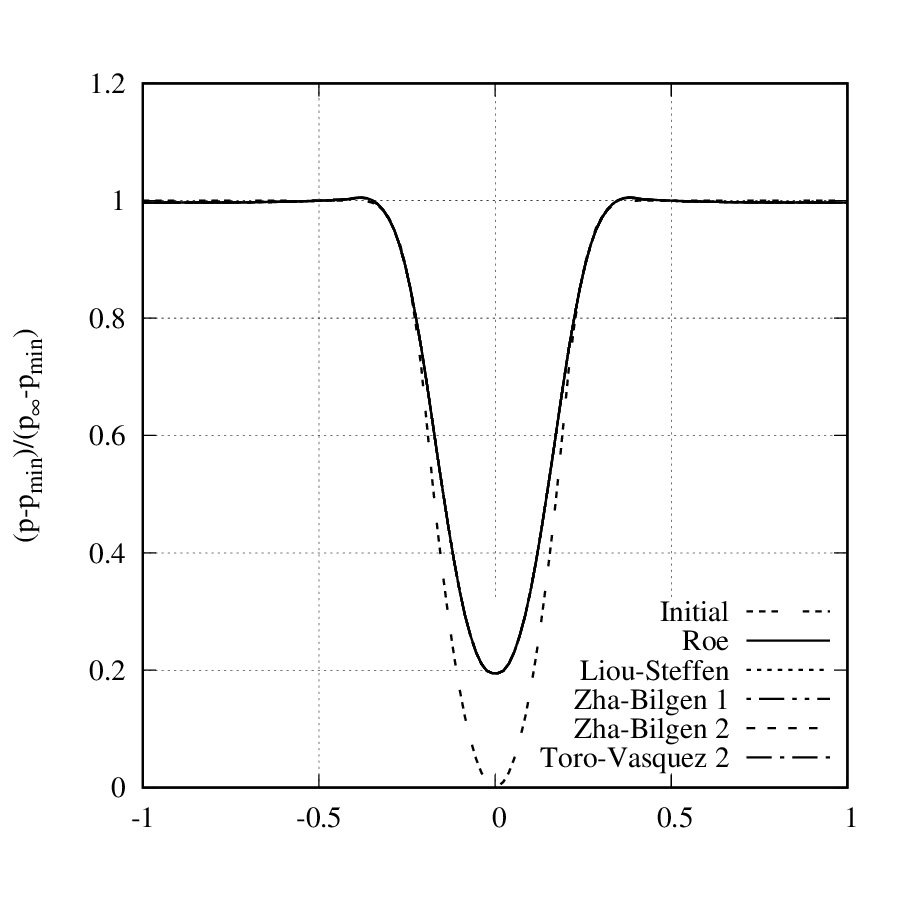}
    \caption{}
    \label{fig:vortex-mixed-pressure}
\end{subfigure}
    \caption{Non-dimensional gauge pressure profiles for a Gresho vortex at $M=0.01$ after $1/4$ rotation using (a) convective diffusion scaling $\uuline{A}^c$ (b) mixed diffusion scaling $\uuline{A}^m$. $p_{\text{min}}$ is the minimum pressure at the centre of the vortex core at $t=0$.}
    \label{fig:vortex-pressure}
\end{figure}

The pressure profile on a line through the centre of the vortex predicted by each of the schemes is shown in figure \ref{fig:vortex-pressure}, where it can be seen that all of the schemes diffuse the vortex to some extent.
With the convective diffusion scaling, the Roe, Liou-Steffen, and second Zha-Bilgen and Toro-Vasquez schemes match once again, the solution from the first Zha-Bilgen scheme is slightly less diffusive, and the first Toro-Vasquez scheme is less diffusive still, in agreement with the limit equation analysis.
The limit equations with the mixed diffusion scaling at the convective limit are the same for all four schemes, and figure \ref{fig:vortex-mixed-pressure} shows that the predicted pressure profiles are all almost identical to each other.
The solution with the first Toro-Vasquez scheme with the mixed diffusion scaling broke down and is not shown.

\begin{figure}
    \centering
    \begin{subfigure}[b]{0.30\textwidth}
        \centering
        \includegraphics[width=0.99\textwidth]{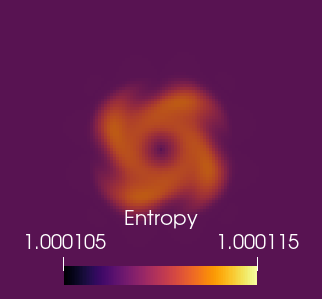}
        \caption{Roe}
        \label{fig:vortex-roe-convective-entropy}
    \end{subfigure}
    \begin{subfigure}[b]{0.30\textwidth}
        \centering
        \includegraphics[width=0.99\textwidth]{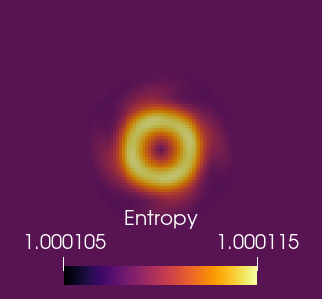}
        \caption{Liou-Steffen}
        \label{fig:vortex-ls-convective-entropy}
    \end{subfigure}
    \vskip\baselineskip
    \begin{subfigure}[b]{0.30\textwidth}
        \centering
        \includegraphics[width=0.99\textwidth]{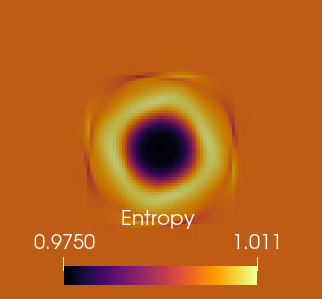}
        \caption{Zha-Bilgen 1}
        \label{fig:vortex-zb-convective-entropy1}
    \end{subfigure}
    \begin{subfigure}[b]{0.30\textwidth}
        \centering
        \includegraphics[width=0.99\textwidth]{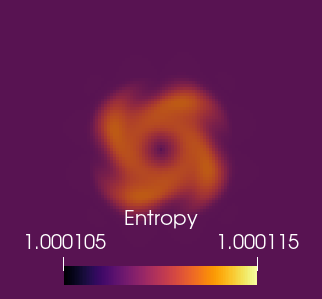}
        \caption{Zha-Bilgen 2}
        \label{fig:vortex-zb-convective-entropy2}
    \end{subfigure}
    \vskip\baselineskip
    \begin{subfigure}[b]{0.30\textwidth}
        \centering
        \includegraphics[width=0.99\textwidth]{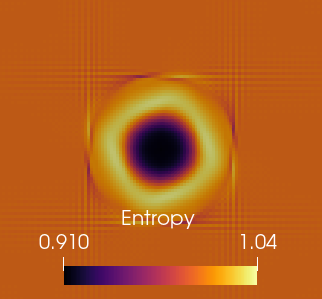}
        \caption{Toro-Vasquez 1}
        \label{fig:vortex-tv-convective-entropy1}
    \end{subfigure}
    \begin{subfigure}[b]{0.30\textwidth}
        \centering
        \includegraphics[width=0.99\textwidth]{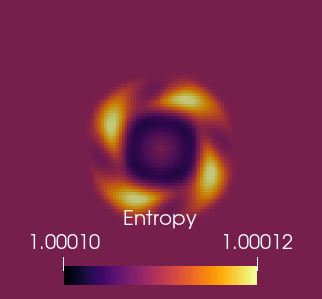}
        \caption{Toro-Vasquez 2}
        \label{fig:vortex-tv-convective-entropy2}
    \end{subfigure}
    \caption{Non-dimensional entropy profiles for a Gresho vortex at $M=0.01$ after $1/4$ rotation using the convective diffusion scaling $\uuline{A}^c$.}
    \label{fig:vortex-convective-entropy}
\end{figure}

\begin{figure}
    \centering
    \begin{subfigure}[b]{0.30\textwidth}
        \centering
        \includegraphics[width=0.99\textwidth]{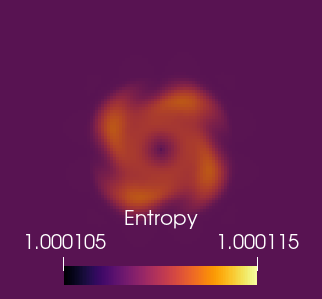}
        \caption{Roe}
        \label{fig:vortex-roe-mixed-entropy}
    \end{subfigure}
    \begin{subfigure}[b]{0.30\textwidth}
        \centering
        \includegraphics[width=0.99\textwidth]{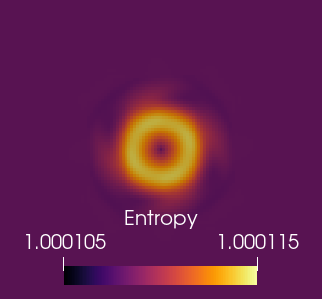}
        \caption{Liou-Steffen}
        \label{fig:vortex-ls-mixed-entropy}
    \end{subfigure}
    \vskip\baselineskip
    \begin{subfigure}[b]{0.30\textwidth}
        \centering
        \includegraphics[width=0.99\textwidth]{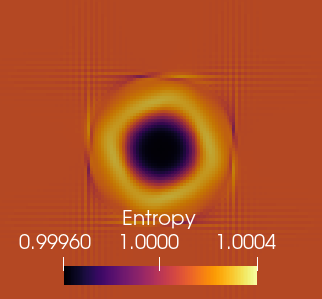}
        \caption{Zha-Bilgen 1}
        \label{fig:vortex-zb-mixed-entropy1}
    \end{subfigure}
    \begin{subfigure}[b]{0.30\textwidth}
        \centering
        \includegraphics[width=0.99\textwidth]{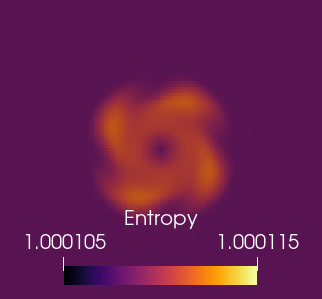}
        \caption{Zha-Bilgen 2}
        \label{fig:vortex-zb-mixed-entropy2}
    \end{subfigure}
    \vskip\baselineskip
    \begin{subfigure}[b]{0.30\textwidth}
        \centering
        \includegraphics[width=0.99\textwidth]{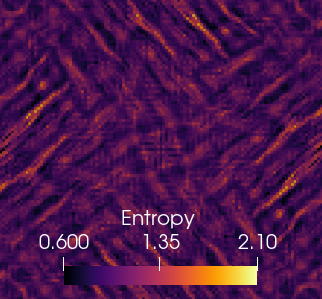}
        \caption{Toro-Vasquez 1}
        \label{fig:vortex-tv-mixed-entropy1}
    \end{subfigure}
    \begin{subfigure}[b]{0.30\textwidth}
        \centering
        \includegraphics[width=0.99\textwidth]{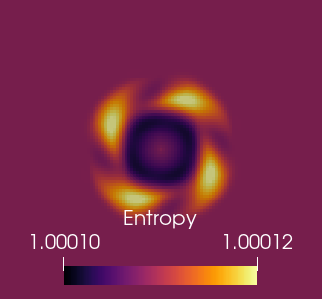}
        \caption{Toro-Vasquez 2}
        \label{fig:vortex-tv-mixed-entropy2}
    \end{subfigure}
    \caption{Non-dimensional entropy profiles for a Gresho vortex at $M=0.01$ after $1/4$ rotation using the mixed diffusion scaling $\uuline{A}^m$.}
    \label{fig:vortex-mixed-entropy}
\end{figure}

The entropy profiles found with the convective and mixed diffusion scalings are shown in figures \ref{fig:vortex-convective-entropy} and \ref{fig:vortex-mixed-entropy} respectively.
The Roe, Liou-Steffen, and second Zha-Bilgen schemes generate very little entropy, and the second Toro-Vasquez generates only a small amount more.
For the Roe and the second Zha-Bilgen and Toro-Vasquez schemes, the entropy generation is predominantly at the four corners where the flow is the most misaligned with the grid.
The Liou-Steffen splitting generates slightly more entropy than the Roe scheme, which is the opposite of the trend seen in the circular cylinder test case.
The profiles in \ref{fig:vortex-roe-convective-entropy} and \ref{fig:vortex-roe-mixed-entropy} are similar to the entropy production modes of an entropy stable Roe scheme shown in figure 11 of \cite{gouasmi_entropy-stable_2022} smeared by a quarter turn, and the entropy production after a single timestep closely resembles the production mode.
The entropy profiles found with the first Zha-Bilgen and Toro-Vasquez forms and the convective scaling have much larger variations, and their shape is consistent with what would be expected from the limit equation (\ref{eq:limit-zb-Ac-convective}) i.e. there appears to be an entropy sink where $\nabla^2p>0$ in the vortex core, and an entropy source where $\nabla^2p<0$ at the edge of the vortex.
The entropy profile for the first Zha-Bilgen form with the mixed diffusion scaling is qualitatively very similar to that with the convective diffusion scaling, except with the variations reduced by around an order of $M$.
However, the solution for the first Toro-Vasquez for with the mixed diffusion scaling has entirely broken down, as can be seen from figure \ref{fig:vortex-tv-mixed-entropy1}, with only very high wavenumber oblique modes remaining.
As with the previous examples, the second forms of Zha-Bilgen and Toro-Vasquez significantly improve the solution compared to the first forms, with the entropy field for the second Zha-Bilgen form almost the same as with the Roe scheme.
The benefits of the second form are especially clear for the Toro-Vasquez scheme with mixed diffusion scaling, preventing the total breakdown that was seen with the first form, with the entropy variations on the same order of magnitude as for the Roe scheme.\\


We have shown three numerical examples which demonstrate the behaviour of the three flux-vector-splittings for both acoustic and convective flow.
Most schemes have the expected property that the mixed diffusion scaling is suitable for both convective and acoustic flow, whereas the convective diffusion scaling will only properly resolve convective flow and will quickly damp out acoustic waves.
However, the first Toro-Vasquez form will not damp out acoustic waves even with the convective scaling.
All three schemes suffer from pressure chequerboard modes with the mixed scaling, but the convective diffusion scaling resolves this issue for all schemes except the first Toro-Vasquez scheme.
The most significant result is the demonstration that the erroneous pressure anti-diffusion terms in the entropy variable limit equations of the Zha-Bilgen and Toro-Vasquez schemes do in fact degrade the solutions found with these schemes.
On the other hand, the Liou-Steffen results almost exactly match the results found with the Roe-type scheme for all cases and with both the convective and mixed diffusion scaling, as predicted by the limit equations analysis.

\section{Conclusions}\label{sec:conclusions}

In this study. we have analysed the form of the artificial diffusion at low Mach number for three convection-pressure flux-vector splittings: Liou-Steffen, Zha-Bilgen, and Toro-Vasquez.
The approximate diffusion form of the Liou-Steffen and Zha-Bilgen splittings bear a close resemblance to that of the Roe scheme, whereas the Toro-Vasquez form does not.
We identified two forms of the energy equation component of the pressure perturbation term for the Zha-Bilgen and Toro-Vasquez splittings - the first form results in diffusion terms with a close correspondence to the original flux-vector splitting, whereas the second results in diffusion terms closer to the Roe and Liou-Steffen forms.

We then transformed the artificial diffusion to the entropy variables to gain more insight into the differences between the splittings.
The limit equations of the Liou-Steffen splitting and the second forms of the Zha-Bilgen and Toro-Vasquez splittings are identical to those of the Roe scheme for both convective and acoustic flow with both the convective and mixed diffusion scalings.
One the other hand, the first forms of the Zha-Bilgen and Toro-Vasquez splittings were found to have erroneous pressure anti-diffusion terms on the entropy equation, which remain in the limit equations with the convective diffusion scaling at both the convective and acoustic limits.
The first form of the Toro-Vasquez splitting also completely lacks the usual pressure diffusion in the pressure equation.

Examining low Mach number convection-pressure flux-vector splitting schemes in the literature, we see that the Liou-Steffen splitting is clearly the most popular, and that almost all modern schemes of all splittings use the mixed diffusion scaling.
All of the Liou-Steffen and Zha-Bilgen schemes reviewed had a form matching the general form assumed in this paper.
The Toro-Vasquez schemes, while mostly matching, apply the velocity perturbation diffusion inconsistently by applying it only to the density and energy equations and not the momentum equations.
All Zha-Bilgen and Toro-Vasquez schemes reviewed used the second form of the diffusion which more closely resembles the Roe and Liou-Steffen forms, instead of the first forms that results in the erroneous entropy generation terms.

Three numerical examples were used to verify the findings of the analysis, showing excellent agreement between prediction and results for all the splittings at both the convective and acoustic limits, including the spurious entropy generation of the first Zha-Bilgen and Toro-Vasquez forms and the significant improvements when using the second forms.

The differences between the three flux-splitting schemes considered here show that the diffusion on the energy equation is an important factor in the design of numerical schemes for low Mach number.
In particular, it is crucial to obtain the correct form of the $\delta U$ diffusion term - the spurious entropy generation of the first Zha-Bilgen and Toro-Vasquez forms is entirely due to this term, irrespective of the form of the $\delta p$ diffusion term.
This is also relevant for future studies of low Mach number numerical schemes because it is not uncommon for the barotropic Euler equations to be used for such studies, which removes the energy equation by enforcing additional constraints on the equation of state.
The barotropic equations are simpler and allow for more in-depth mathematical analysis, but clearly care must be taken when extending the findings of such analysis to the full Euler equations where the energy equation must also be considered.
Future extensions of convection-pressure flux-vector splittings could include alternative methods for ensuring correct resolution of the entropy field, the effect of including $\delta U$ in deciding the upwind direction, and other forms for the $\delta(pU)$ term in the pressure perturbation.

%
%

%
%
%
%
%
%
%

\section*{Acknowledgements}
The authors gratefully acknowledge support from the EPSRC Center for Doctoral Training in Gas Turbine Aerodynamics and Rolls-Royce plc under grant number EP/L015943/1.

\printbibliography







\end{document}